\documentclass[fleqn]{aa}
\usepackage{mathtools}
\usepackage{txfonts}
\usepackage{natbib}
\usepackage{graphicx}
\usepackage{epstopdf}
\usepackage{epsfig}
\usepackage{pgf}
\usepackage{amsmath}

\usepackage{soul}
\definecolor{lightgreen}{HTML}{B7F774}
\sethlcolor{yellow}

\providecommand{\mc}[1]{\multicolumn{1}{c}{#1}}

\begin{document}

\title{Multiband {\em RadioAstron} space VLBI imaging of the jet in quasar S5\,0836$+$710}

\author{L.~Vega-Garc\'ia\inst{1}
        \and
        A.~P.~Lobanov\inst{1,2}
        \and
        M.~Perucho\inst{3,4}
        \and 
        G.~Bruni\inst{5,1}
        \and
        E.~Ros\inst{1}
        \and
        J.~M.~Anderson\inst{6,7,1}
        \and
        I.~Agudo\inst{8}
        \and
        R.~Davis\inst{9}$^\ddag$
        \and
        J.~L.~G\'omez\inst{8}
        \and
        Y.~Y.~Kovalev\inst{10,11,1}
        \and
        T.~P.~Krichbaum\inst{1}
        \and
        M.~Lisakov\inst{1,10}
        \and
        T.~Savolainen\inst{12,13,1}
        \and
        F.~K.~Schinzel\inst{14,15}
        \and
        J.~A.~Zensus\inst{1}
}

    \institute{Max-Planck-Institut f\"ur Radioastronomie,
              Auf dem H\"ugel 69, D-53121 Bonn, Germany
    \and
    Institut f\"ur Experimentalphysik, Universit\"at Hamburg, 
    Luruper Chaussee 149, D-22761 Hamburg, Germany
    \and
    Departament d'Astronomia i Astrof\'isica, Universitat de Val\`encia, C/ Dr. Moliner, 50, E-46100 Burjassot, Val\`encia, Spain 
    \and
    Observatori Astron\`omic, Universitat de Val\`encia, C/ Catedr\`atic Beltr\'an 2, E-46091 Paterna , Val\`encia, Spain
    \and 
    INAF - Istituto di Astrofisica e Planetologia Spaziali, via Fosso del Cavaliere 100, I-00133 Rome, Italy
    \and
    Deutsches GeoForschungsZentrum, Telegrafenberg, D-14473, Potsdam, Germany
    \and
    Technical University of Berlin, Straße des 17. Juni 135, 10623 Berlin, Germany
    \and
    Instituto de Astrof\'isica de Andaluc\'ia, CSIC, Glorieta de la Astronom\'{\i}a s/n, E-18008, Granada, Spain
    \and
    Jodrell Bank Observatory (JBO), The University of Manchester, Macclesfield SK11 9DL, United Kingdom
    \and
    Astro Space Center of Lebedev Physical Institute, Profsoyuznaya 84/32, 117997 Moscow, Russia
    \and
    Moscow Institute of Physics and Technology, Dolgoprudny, Institutsky per., 9, Moscow region, 141700, Russia
    \and
    Aalto University Department of Electronics and Nanoengineering, PL 15500, FI-00076 Aalto, Finland
    \and
    Aalto University Mets\"ahovi Radio Observatory, FIN-02540 Kylm\"al\"a, Finland
    \and
    National Radio Astronomy Observatory, P.O. Box O, Socorro, NM 87801 USA
    \and
    Department of Physics and Astronomy, University of New Mexico, Albuquerque NM, 87131, USA
    }

\authorrunning{Vega-Garc\'ia et al.}
\titlerunning{Multiband {\em RadioAstron} imaging of S5\,0836$+$710}

  \date{}
 
  \abstract
  {Detailed studies of relativistic jets in active galactic nuclei
    (AGN) require high-fidelity imaging at the highest possible
    resolution. This can be achieved using very long baseline
    interferometry (VLBI) at radio frequencies, combining worldwide
    (\textup{global}) VLBI arrays of radio telescopes with a space-borne
    antenna on board a satellite.}
  {We present multiwavelength images made of the radio emission in the
    powerful quasar S5\,0836$+$710, obtained using a global VLBI array
    and the antenna {\em Spektr-R} of the {\em RadioAstron} mission of the
    Russian Space Agency, with the goal of studying the internal
    structure and physics of the relativistic jet in this object.}
  {The {\em RadioAstron} observations at wavelengths of 18\,cm, 6\,cm,
    and 1.3\,cm are part of the Key Science Program for imaging radio
    emission in strong AGN. The internal structure of the jet is
    studied by analyzing transverse intensity profiles and modeling
    the structural patterns developing in the flow.}
  {The {\em RadioAstron} images reveal a wealth of structural detail
    in the jet of S5\,0836+710 on angular scales ranging from
    0.02\,mas to 200\,mas. Brightness temperatures in excess of
    $10^{13}$\,K are measured in the jet, requiring Doppler factors of
    $\ge 100$ for reconciling them with the inverse Compton
    limit. Several oscillatory patterns are identified in the
    ridge line of the jet and can be explained in terms of
    the Kelvin-Helmholtz (KH) instability. The oscillatory patterns are
    interpreted as the surface and body wavelengths of the helical
    mode of the KH instability. The interpretation provides estimates of
    the jet Mach number and of the ratio of the jet to the ambient density, which are
    found to be $M_\mathrm{j}\approx 12$ and $\eta\approx 0.33$. The
ratio of the    jet to the ambient density should be conservatively
    considered an upper limit because its estimate
    relies on approximations.}
{}

   \maketitle
%

\section{Introduction}
 
Very long baseline interferometry (VLBI) that combines ground-based radio
telescopes with an orbiting antenna to engage in \textup{}space VLBI observations is an established tool for reaching an unprecedentedly
high angular resolution of astronomical measurements (see
\citealt{arsentev+1982,levy+1986,hirabayashi+2000} for overviews of
earlier space VLBI programs). The ongoing international space VLBI
mission {\em RadioAstron} \citep{kardashev+2013} led by the Astro
Space Center (ASC, Moscow, Russia) reaches a resolution of $\sim 10$
 $\mu$as at a frequency of 22\,GHz. The space segment
of {\em RadioAstron} employs a 10-meter antenna (Space Radio
Telescope, SRT) on board the satellite {\em Spektr-R,} which launched on
18 July 2011.  {\em Spektr-R} has a highly elliptical orbit with a
period of 8 to 9 days and an apogee reaching up to 350\,000\,km.  The
SRT operates at frequencies 0.32, 1.6, 5, and 22\,GHz
\citep{kardashev+2013}.  The SRT delivers data in both left (LCP) and
right (RCP) circular polarization at 0.32, 1.6, and 22\,GHz. At
5\,GHz, the SRT provides only LCP data.  The {\em
  RadioAstron} observations are correlated at three different facilities: the ASC {\em RadioAstron} correlator in Moscow
\citep{likhachev+2017}, the JIVE\footnote{Joint Institute for VLBI in
  Europe, Dwingeloo, The Netherlands.} correlator in Dwingeloo
\citep{kettenis2010}, and the DiFX correlator
\citep{deller+2007,deller+2011} of the
MPIfR\footnote{Max-Planck-Institut f\"ur Radioastronomie, Bonn,
  Germany.\\ $^\ddag$ -- deceased.} in Bonn, which uses a dedicated
DiFX correlator version built to process data from space-borne antennas
\citep{bruni+2015,bruni+2016}.

Several {\em RadioAstron} campaigns are performed within three Key
Science Programs (KSP) on imaging compact radio emission in radio-loud
active galactic nuclei (AGN). {\em RadioAstron} observations of the
high-redshift quasar S5\,0836$+$710 presented in this paper constitute
part of the KSP observations of the parsec-scale structure of powerful
AGN. The main goal of this KSP is to study the innermost regions of
jets for the most prominent blazars, particularly, to study the
evolution of shocks and plasma instabilities in the flow.

The radio source S5\,0836$+$710 (4C\,$+$71.07; J0841$+$7053) is a
powerful low-polarization quasar (LPQ) located at a redshift of 2.17
\citep{osmer+1994}, which corresponds to the luminosity distance of
16.9\,Gpc and to a linear scale of 8.4\,pc/mas, assuming the standard
$\Lambda$CDM cosmology \citep{planck2015}.  It has a long, one-sided
jet at parsec and kiloparsec scales.  The large-scale structure of
S5\,0836$+$710 was previously imaged with the VLA\footnote{Karl
  G. Jansky Very Large Array of the National Radio Astronomy
  Observatory, Socorro, NM, USA} and MERLIN\footnote{Multi-Element
  Radio Linked Interferometer Network of the Jodrell Bank Observatory,
  UK} at distances larger than 1 arcsecond
\citep{hummel+1992,perucho+2012b}.  The optical spectrum of
S5\,0836$+$710 shows broad lines \citep{torrealba+2012}. In the
parsec-scale jet of the source, apparent speeds of up to $21\,c$ have
been reported \citep{lister+2013}. The estimated jet viewing angle is
$\theta \approx 3.2 ^\circ$ \citep{otterbein+1998}.

The source morphology suggests a likely presence of plasma instability
in the flow.  \cite{krichbaum+1990} observed the source with VLBI at
326\,MHz and 5\,GHz and identified several bends in the flow that can
be associated with the growth of instability \citep[see][for a recent
review]{perucho2019}.  High-resolution images obtained with VLBI Space
Observatory Program \citep[VSOP;][]{hirabayashi+2000} at 1.6\,GHz and
5\,GHz \citep{lobanov+1998} revealed a positional offset (``core
shift'') of $\approx 1.0$\,mas measured for the jet base observed at
these two frequencies \citep{lobanov+2006}, which is most likely
caused by the opacity gradient along the jet due to the synchrotron
self-absorption \citep{koenigl1981}. The measured 1.6--5\,GHz core
shift can be used to estimate basic physical properties of the jet
base imaged at 5\,GHz \citep[see][]{lobanov1998}, including the
magnetic field, $B_\mathrm{core,5GHz} \approx 20$\,mG, and the
distance $r_\mathrm{core,5GHz} \approx 70$\,pc of that region from
the true jet origin. Extrapolation of these estimates to the highest
{\em RadioAstron} observing frequency of 22\,GHz gives
$B_\mathrm{core,22GHz} \approx 100$\,mG and $r_\mathrm{core,22GHz}
\approx 15$\,pc \citep[or $\approx 1.6 \times 10^5$ gravitational
radii, $R_\mathrm{g}$, using the estimated black hole mass of $2\times
10^9\,\mathrm{M}_{\odot}$;][] {tavecchio+2000}.  This places the
observed jet base at distances that are at least an order of
magnitude larger than the typical Poynting flux dissipation scale,
$r_\mathrm{PFD} \sim 10^4\,R_\mathrm{g}$, expected for magnetized
flows \citep{komissarov+2007,lyubarsky2010,mertens+2016,mizuno+2016}. 
Even in the most extreme scenarios with $r_\mathrm{PFD} \lesssim
10^6\,R_\mathrm{g}$ (\cite{nakamura+2004,vlahakis+2004}), the regions of
the jet imaged by {\em RadioAstron} in S5\,0836$+$710 should lie
outside the flow zone that is dominated by Poynting flux. The jet in these
images can therefore be considered kinetically dominated everywhere
downstream of its observed base. This consideration leaves
Kelvin-Helmholtz (KH) instability as the most likely process
responsible for the observed jet structure probed by the {\em
  RadioAstron} observations of S5\,0836$+$710. The current-driven instability may play a
  role in initially stabilizing the jet against the KH instability
  and affecting the wavelengths and growth rates of the latter
  \citep[see][]{mizuno+2012,mizuno+2014}, but it would require
  observations at frequencies higher than 22\,GHz to uncover the
  relation between the two types of instability in S5\,0836$+$710. 

The jet ridge line measured in the VSOP and ground VLBI images
can indeed be reconciled with the presence of KH instability in
the flow
\citep{lobanov+1998,perucho+2007b,perucho+2012a,vegagarcia2019}. The
oscillations observed in the ridge line with respect to the overall jet
direction were associated with the helical, $H_\mathrm{s}$, and
elliptical, $E_\mathrm{s}$, surface modes of Kelvin-Helmholtz
instability \citep{hardee+1997,hardee2000}, which introduce ellipticity in
the cross-section of the flow ($E_\mathrm{s}$ mode) and force the
entire flow to oscillate around its main propagation direction
($H_\mathrm{s}$ mode). Investigations of more intricate body modes of
the instability that affects the jet interior \citep[see][]{hardee2000}
have not been feasible so far for S5\,0836$+$710, owing to
insufficient transverse resolution of the jet structure in VLBI images
made for this object.

Observations of the jet made with
MERLIN{} at arcsecond scales showed an emission gap between
0.2\,$\arcsec$ and 1.0\,$\arcsec$ and a large-scale structure beyond
1\,$\arcsec$.  This was explained by the acceleration and expansion of
the jet and its disruption due to a helical instability
\citep{perucho+2012b}.  The presence of shocks has been suggested in
the jet on scales up to $\sim 0.5\,$kpc, which were revealed by multiple regions with
flatter spectrum that are separated by $\simeq 5~{\rm mas}$
\citep{lobanov+2006}.

The multiband {\em RadioAstron} observations of S5\,0836$+$710
presented here are described in Sect.~\ref{sc:method}.  The resulting
images are introduced in Sect.~\ref{sc:image}, the interpretation of
the asymmetric structure of the jet is discussed in Sect.~\ref{sc:asy}, and
the large-scale jet structure is analyzed in
Sect.~\ref{sc:ridgeLineOscillations}. The results are summarized and
discussed in a broader context in Sect.~\ref{sc:summary}.

\section{{\em RadioAstron} observations of S5\,0836$+$710}
\label{sc:method}

\begin{figure}[t!]
\centerline{\includegraphics[width=0.3\textwidth,angle=-90,trim=27mm 30mm 10mm 35mm,clip=true]{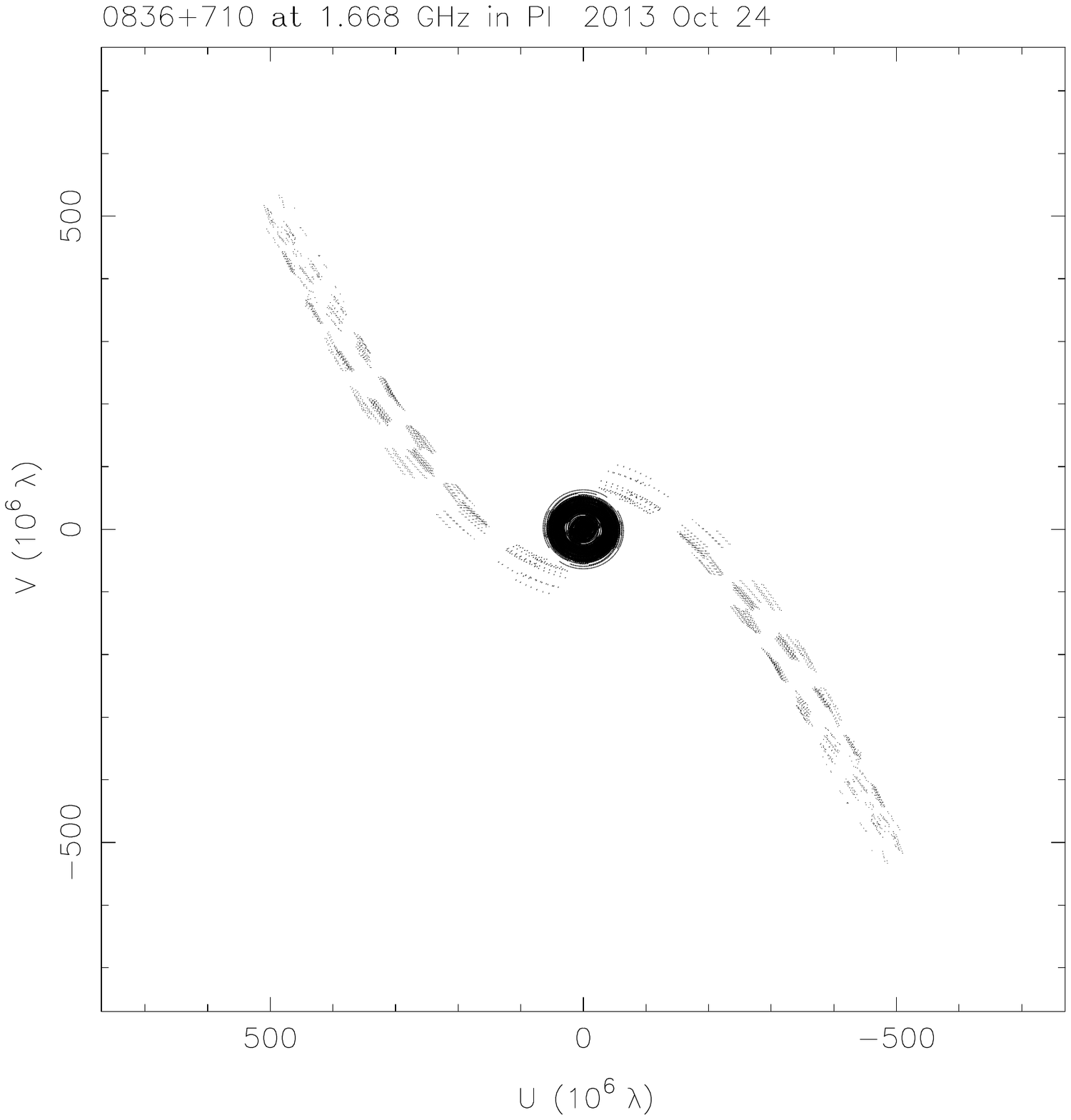}}
\centerline{\includegraphics[width=0.3\textwidth,angle=-90,trim=27mm 30mm 10mm 35mm,clip=true]{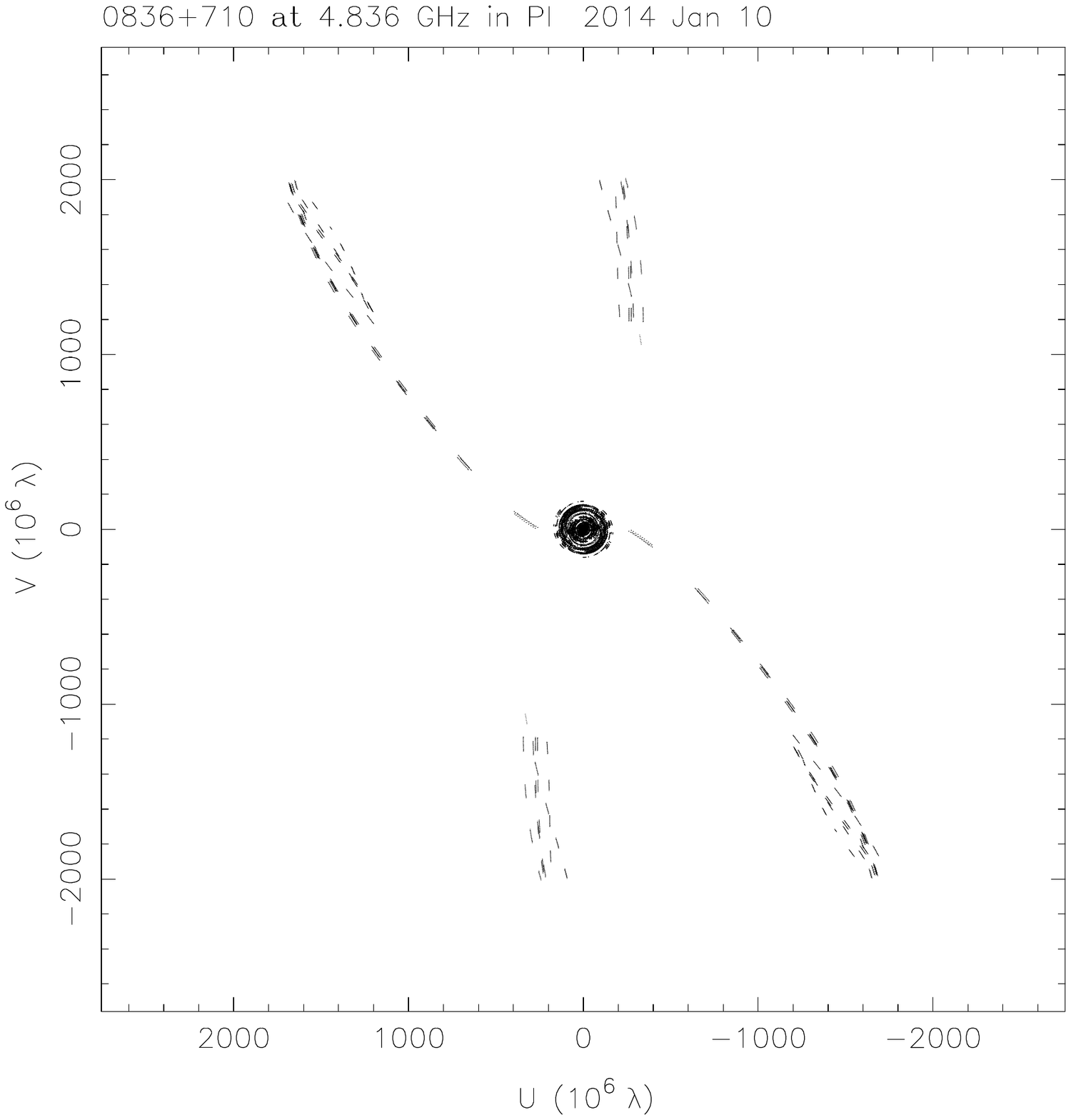}}
\centerline{\includegraphics[width=0.3\textwidth,angle=-90,trim=27mm 30mm 10mm 35mm,clip=true]{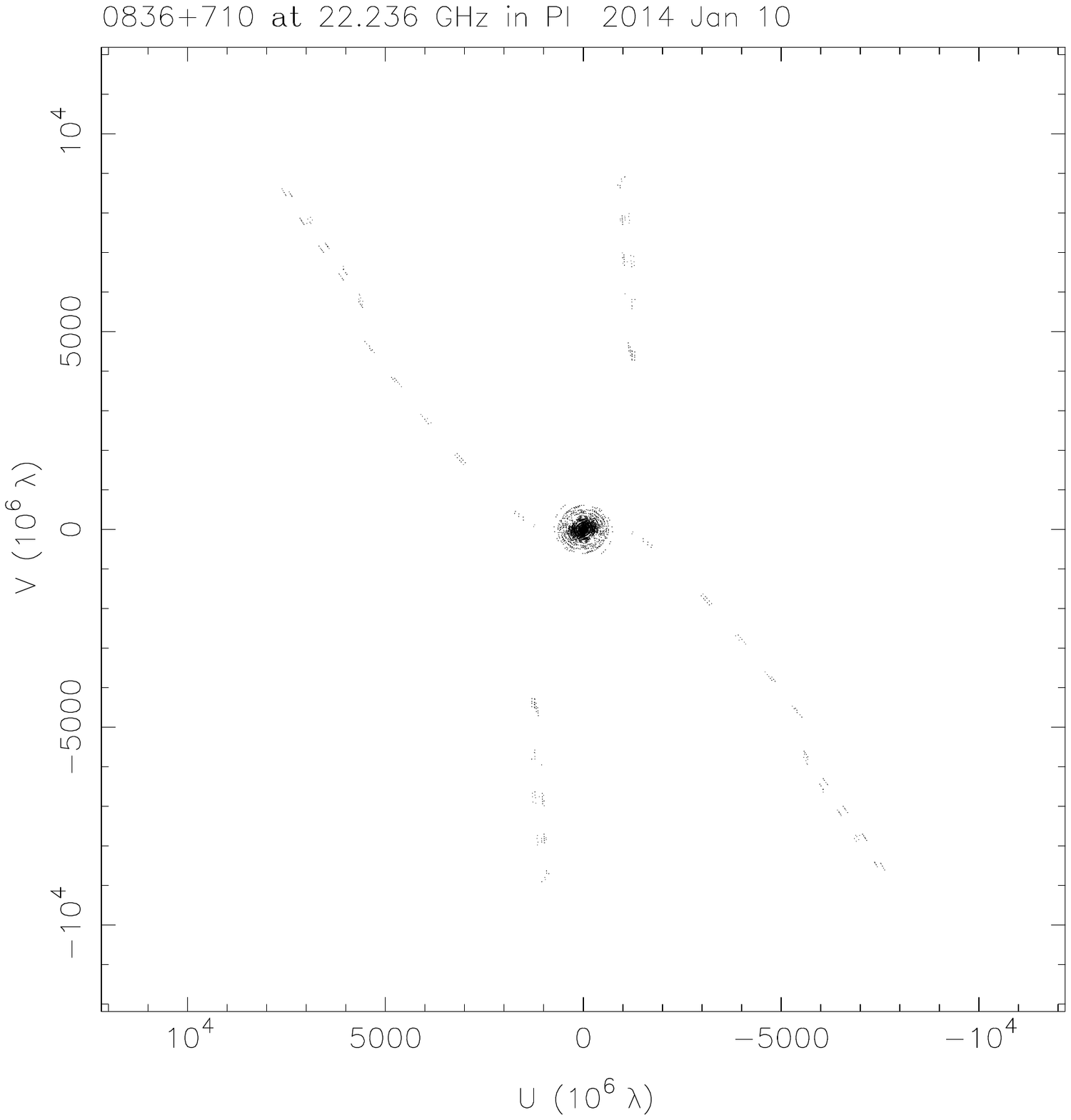}}
\caption{Coverage of the Fourier domain ({\em uv} coverage) of the
  {\em RadioAstron} observations of S5\,0836$+$710 at 1.6\,GHz (top),
  5\,GHz (middle), and 22\,GHz (bottom) plotted in units of M$\lambda$
  at each respective frequency. Central concentrations correspond
  to baselines between the ground antennas ({\em ground--ground baselines}). Long arcs represent
  baselines between the ground antennas and the SRT ({\em ground--space
  baselines}).}
\label{fg:0836-uvplot}
\end{figure}

{\em RadioAstron} imaging observations of S5\,0836$+$710 were
performed at two epochs: on 24 October 2013 at L band (1.6\,GHz,
wavelength $\lambda$$=$18\,cm) and on 10--11 January 2014 at C band
(5\,GHz, $\lambda$$=$6\,cm) and K band (22\,GHz, $\lambda$$=$1.3\,cm),
combining data recorded at the SRT with data provided by a ground
array of radio telescopes. At each epoch the source was observed for
about 18 hours, starting from a perigee passage of the SRT, in order
to ensure a continuous coverage of Fourier domain ({\em uv} coverage)
by the ground--space baseline data. The resulting {\em uv} coverage of
the {\em RadioAstron} observations are shown in
Fig.~\ref{fg:0836-uvplot}.

\begin{table*}[t!]
\caption{Radio telescopes participating in the {\em RadioAstron} observations of S5\,0836$+$710}
\label{tb:telescopes}
\begin{center}
\begin{tabular}{rrlrrrrr}\hline\hline
Telescope      &    &   & \multicolumn{1}{c}{L-band} & \multicolumn{1}{c}{C-band} & \multicolumn{1}{c}{K-band}\\
      &    & \mc{$D$} & SEFD   & SEFD   &  SEFD    \\
               & Code   & [m] & [Jy]   &  [Jy]    & [Jy]    \\\hline
{\em Spektr-R} SRT (RU) & RA & ~~10       & 2840   & 9500   & 30000  \\
Brewster (USA)          & BR & ~~25       & 282    & 207    & 608    \\
Fort Davis (USA)        & FD & ~~25       & ...    & 186    & 480    \\
Hancock (USA)           & HN & ~~25       & 324    & 261    & 850    \\
Kitt Peak (USA)         & KP & ~~25       & 271    & 196    & 452    \\
Los Álamos (USA)       & LA & ~~25       & ...    & 202    & 370    \\
Mauna Kea (USA)         & MK & ~~25       & 418    & 238    & 380    \\
North Liberty (USA)     & NL & ~~25       & 324    & 210    & 711    \\
Owens Valley (USA)      & OV & ~~25       & 354    & 221    & 594    \\
Pie Town (USA)          & PT & ~~25       & 281    & 202    & 474    \\
St. Croix (USA)         & SC & ~~25       & 304    & 233    & 1045   \\
Green Bank (USA)        & GB & 100$^\dag$  & 10     & 10     & 20     \\
Effelsberg (DE)         & EF & 100        & 19     & 20     & 90     \\
Jodrell Bank (UK)       & JB & ~~76       & 65     & 80     & ...    \\
WSRT (NL)               & WB & ~~66$^\dag$ & 40     & 120    & ...    \\
Torun (PL)              & TR & ~~32       & 300    & ...    & ...    \\
Badary (RU)             & BD & ~~32       & 330    & ...    & ...    \\
Svetloe (RU)            & SV & ~~32       & 360    & ...    & ...    \\
Urumqi (CH)             & UR & ~~25       & 300    & ...    & ...    \\
Shanghai (CH)           & SH & ~~65       & 670    & ...    & ...    \\
Noto (IT)               & NT & ~~32       & 784    & ...    & ...    \\
Medicina (IT)           & MC & ~~32       & 700    & ...    & ...    \\
Onsala (SE)             & ON & ~~25       & 320    & ...    & ...    \\
Yebes (ES)              & YS & ~~40       & ...    & ...    & 200    \\
Zelenchukskaya (RU)     & ZC & ~~32       & 300    & ...    & ...    \\
Kalyazin (RU)           & KL & ~~46       & 138    & 147    & ...    \\ \hline
\end{tabular}
\end{center} 
{\small {\bf Column designation:} $D:$ antenna diameter
  ($\dag$ -- equivalent diameter). SEFD: system equivalent flux
  density (system noise) for the frequencies at which each individual
  antenna participated in the observations.}
\end{table*} 

\begin{table*}[t!]
\caption{Summary of the observations}
\label{tb:obspar}
\begin{center}
\begin{tabular}{lllcccrcc}\hline\hline
\,{\em RadioAstron} & global VLBI  &  Date & Observing time (UT) & $\nu_\mathrm{obs}$ & Polarization & on/off cycle\\
project code        & project code &       &                     &  [GHz]              &              & [min]       \\ \hline
raks05a & GL038A   & 2013/10/24 & 22:00–14:30  & 1.660 & LCP, RCP & 40/70\\ 
raks05b & GL038B & 2014/01/10 & 04:00–12:00  & 4.828\,/\,22.228     & LCP  & 30/95\\ 
raks05b & GL038C & 2014/01/10--11 & 16:00–10:00  & 4.828\,/\,22.228    & LCP      & 30/95\\ \hline 
\end{tabular}
\end{center}
{\small {\bf Notes:}~Polarization channels. LCP: left circularly polarized. RCP: right circularly polarized. $\nu_\mathrm{obs}$ : central frequency in a respective observing band.}
\end{table*}

The L-band data were recorded at all participating antennas in
dual-circular polarisation, comprising the LCP and RCP channels. The
C- and K-band data were recorded at the SRT simultaneously in the
mixed C/K mode, with the SRT recording in the LCP mode simultaneously
at both frequencies (with two IF bands per frequency). At each
frequency, the SRT recording was supported by a subset of the ground
telescopes (see Table~\ref{tb:telescopes} for details of the frequency
allocation) recording in the dual-circular polarization mode at the
selected frequency band.

General technical parameters of the telescopes participating in the
different observations are given in Table~\ref{tb:telescopes}. The
individual observations and observational setups are summarized in
Table~\ref{tb:obspar}.  The data were recorded at a rate of
128\,megabits per second (Mbps) split into two intermediate-frequency
(IF) bands, resulting in a total bandwidth of 2x16\,MHz for each
circular polarization channel \citep{andreyanov+2014}.  The tracking
stations at Puschino \citep{andreyanov+2014} and Green Bank
\citep{ford+2014} received and recorded the SRT data and telemetry.

The {\em RadioAstron} observations were complemented with simultaneous
VLBA observations at 15\,GHz ($\lambda=2$\,cm) and 43\,GHz
($\lambda=0.7$\,cm) made during gaps between the SRT scans, which are
required for cooling of the onboard high-gain antenna hardware
(consult Table~\ref{tb:obspar} for the duration of the on/off cycles
at the SRT). The respective {\em uv} coverage and images obtained
from these observations are shown in Appendix~\ref{ap:images}
(Figs.~\ref{fg:0836-43-gridge}, \ref{fg:0836-15-gridge},
\ref{fg:0836-15-uvplot}, and \ref{fg:0836-43-uvplot}).

After each of the observing runs, the respective data from all
participating telescopes were transferred to the VLBI correlator of
the MPIfR.  For the correlation and further calibration of {\em
  RadioAstron} data, information on the orbital position,
velocity, and acceleration of the space antenna is introduced in the correlator model.  The
orbit and the momentary state vector of the SRT were reconstructed by
the {\em RadioAstron} ballistic team by combining information
of the radiometric range and radial velocity, measured in Bear Lakes
and Ussurijsk (Russia), Doppler-tracking was performed at both tracking
stations, and measurements of the sky position of the satellite
were obtained from laser ranging and optical astrometry
\citep{khartov+2014,zakhvatkin+2014}. 

\subsection{Data correlation}

The data were processed using the DiFX correlator of the MPIfR at Bonn
upgraded for {\em RadioAstron} data correlation \citep{bruni+2016}.
Fringe-searching for the SRT was performed separately for each individual
observing scan, in order to optimize the centering of the correlation
window, and to minimize residual acceleration terms \citep[see][for a
more detailed discussion]{lobanov+2015}.  An initial search window
with 1024 spectral channels per IF band and 0.1\,s integration time was
used.  Whenever feasible, the delay and rate solutions for 
scans with no fringe detections were interpolated from the adjacent scans
with successful fringe detections.

In the L-band observations, about half of the {\em RadioAstron} scans
yielded fringes at the correlation stage, corresponding to the initial
$\sim$50 minutes of observations, performed with the Green Bank
tracking station, and a further $\sim$3 hours (distributed from 00 UT
to 07 UT) with the Puschino tracking station. The final $\sim$3 hours
(from 08 to 14 UT) of observing time were tracked again by Green Bank,
and initial best-guess delays and rates were extrapolated from the
earlier scans. The detection significance progressively diminished
with increasing baseline length, dropping below a signal-to-noise
ratio (S/N) of 10 at baselines longer than six Earth diameters,
progressively resolving out the source structure at longer space
baselines.

For the C- and K-band observations, only 4 scans out of 30
gave fringes on space baselines with an S/N>10 at the correlator
stage. These scans corresponded to the perigee part of the orbit
($\sim$40 minutes, from 16 UT to 18 UT).  Similar to the approach
employed for the L-band data, the initial delay and rate for the
remaining {\em RadioAstron} scans were extrapolated ($\sim$2.5 hours
from 4 UT to 11 UT, and $\sim$4 hours from 18 UT to 08 UT) and
supplied to the correlator. The final correlation of the data was made
with 64 spectral channels per IF and 0.5\,s of integration time for
the L band, and 32 spectral channels per IF and 0.5\,s of integration
time for the C and K bands. These extrapolated solutions were
subsequently refined and improved during fringe fitting performed as
part of post-processing of the correlated data.

\begin{figure}[t!]
\centerline{\includegraphics[width=0.5\textwidth,angle=0,trim=5 25 0 25,clip=true]{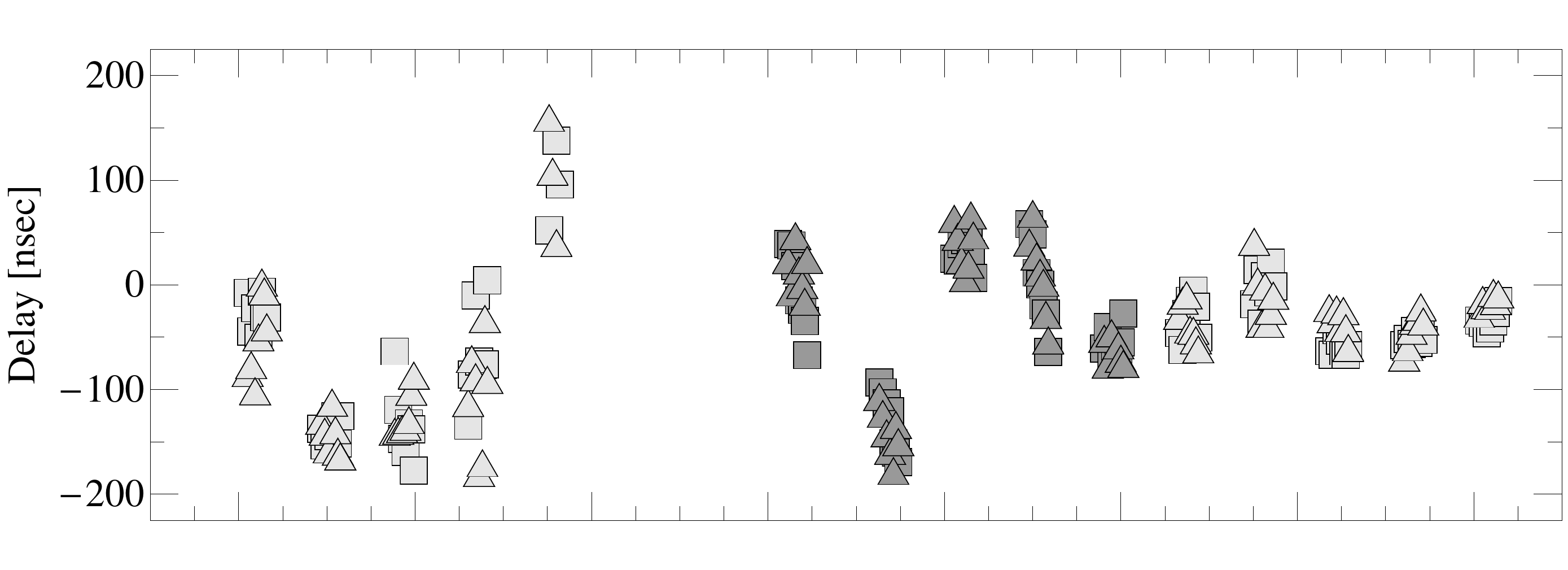}}
\centerline{\includegraphics[width=0.5\textwidth,angle=0,trim=5 00 0 25,clip=true]{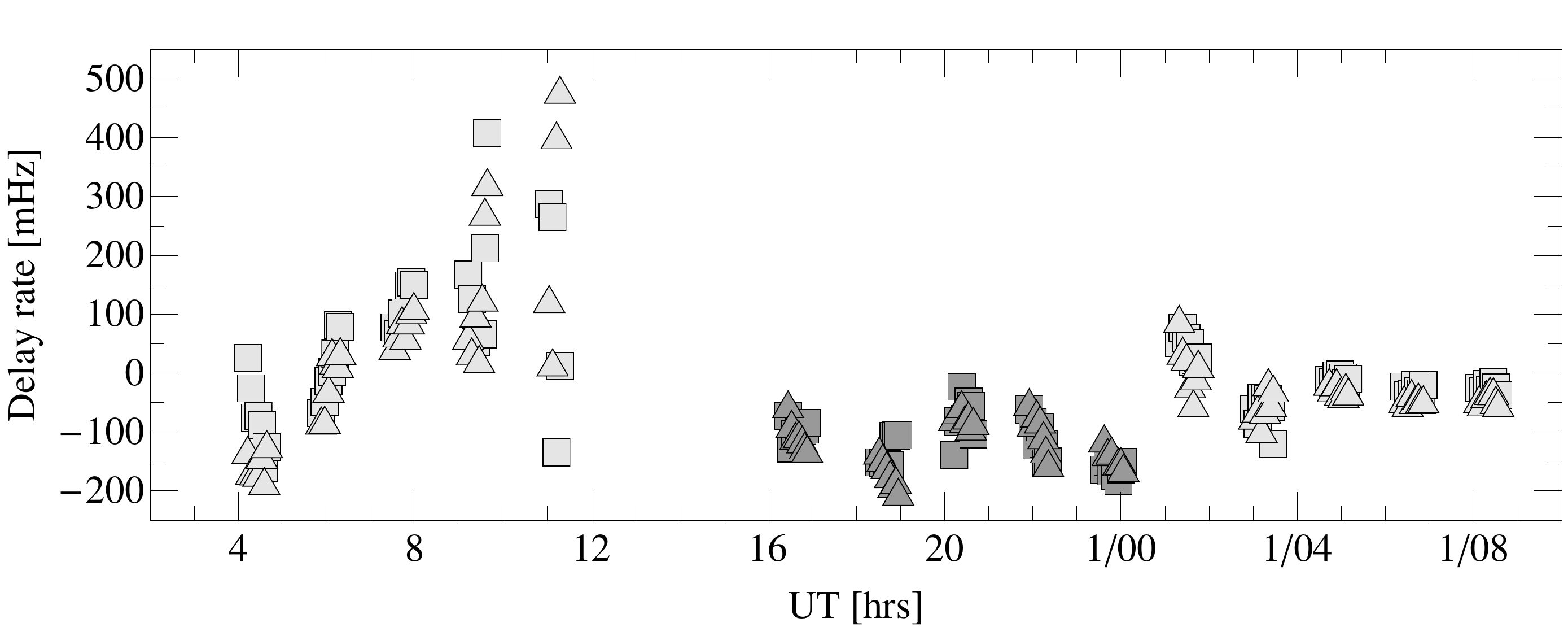}}
\caption{Residual fringe delay (top) and delay rate (bottom) solutions
  at 5\,GHz, obtained for the SRT from the data recorded at the
  tracking stations in Green Bank (light gray, UT 04--12,\,10 January and
  UT 01--08, 11 January) and Puschino (dark gray, UT 16--24,\,10 January). The
  squares and triangles denote the solutions for the first and second
  IF band, respectively. The time variations of the residual delays
  and the delay rates likely reflect the effect of the accelerated
  motion of the SRT, which was not accounted for during the fringe
fitting.}
\label{fg:fringe-sols} 
\end{figure}

\subsection{Post-correlation data reduction}

The correlated data were reduced using AIPS\footnote{Astronomical
  Image Processing Software of the National Radio Astronomy
  Observatory, USA} for the initial calibration and \textit{Difmap}
\citep{shepherd1997,shepherd2011} for imaging.
The \textup{}a priori amplitude calibration was applied using nominal
values for the antenna gains and system temperature measurements made
at each antenna during the observation. For the SRT, the sensitivity
parameters measured in 2011--2013
\citep{kovalev+2014} were used. The data were edited using the
flagging information from each station log.  Parallactic angle
correction was applied to the ground-array antennas to
account for the axis rotation of the dual-polarization antenna feeds
with respect to the target source.

\subsubsection{Fringe fitting}

The data were fringe fit in two steps, first only for the ground
array, and then also including the space baselines, with stacked
solutions for the ground baselines (baseline stacking) used to improve
the detectability of space baseline fringes.  For the observations at
1.6\,GHz, the solution interval for the ground- and space-array fringe
fitting was 4\,min and the cutoff $\mathrm{S/N}=3$ was
set. Effelsberg was used as reference antenna for the first fringe
fitting and Green Bank for the second.  For the 5\,GHz data, the
solution interval was 1\,min and 4\,min for the ground and space
arrays, respectively. The S/N cutoff was set to $\mathrm{S/N} = 4.3$
for the ground VLBI data and $\mathrm{S/N} = 3$ for the space
baselines. Effelsberg was the reference antenna for all baselines.
For the observations at 22\,GHz, the solution intervals for the ground
and space arrays were 1\,min and 5\,min, respectively. The S/N
cutoffs and the reference antenna were the same as used for the data
at 5\,GHz. For the SRT, a two-way running average smoothing over a time
interval of 30 minutes was applied to the solutions obtained at 5 and
22\,GHz. Figure~\ref{fg:fringe-sols} shows the resulting fringe delay
and delay rate solutions obtained for the SRT at 5\,GHz. The solutions
are consistent between the two IF bands, and the likely effect of the
satellite acceleration is reflected both in the time
variations of the fringe delays and in the respective delay rates. The
overall behavior of the fringe solutions is in agreement with the
expected accuracy of the orbital position and velocity determination
\citep{kardashev+2013,duev+2015,lobanov+2015}.

\subsubsection{Bandpass calibration}

For the 1.6\,GHz data, corrections for the individual receiver
bandpasses were introduced. For this calibration, the autocorrelated data
were used and phases and amplitudes were corrected.  The correction was
applied for the whole time interval, and Effelsberg was used as the
reference antenna.  For the 5\,GHz and 22\,GHz data, the attempted bandpass
corrections did not show any improvement and were not implemented.

\section{{\em RadioAstron} images of S5\,0836$+$710}
\label{sc:image}

After the a priori{\em } amplitude calibration, fringe fitting, and
bandpass calibration, the data were averaged in time over 10\,s and in
frequency over all channels in a given IF band and exported to
\textit{Difmap} for imaging and further analysis.  The data were
imaged using the CLEAN hybrid imaging algorithm \citep{hogbom1974} and
self-calibration, with amplitude adjustments limited to a single gain
correction applied to each antenna over the entire duration of the
respective observation. The natural weighting (increasing the
sensitivity to extended emission) was applied to the visibility data
for obtaining the ground array images and the uniform weighting
(maximizing the image resolution) was used to produce the space
VLBI images \citep[see][for a detailed discussion of the data
  weighting in VLBI]{briggs+1999}. Radial distributions of the final
self-calibrated visibility amplitudes and the respective CLEAN models
from the space VLBI images are plotted in Fig.~\ref{fg:0836-radplot}
of Appendix~\ref{ap:images}.  The figures show that structure is
detected up to the longest ground-space baselines and that the CLEAN
models agree well with the data.

\begin{figure}[t!]
  \centering
  \includegraphics[width=0.43\textwidth]{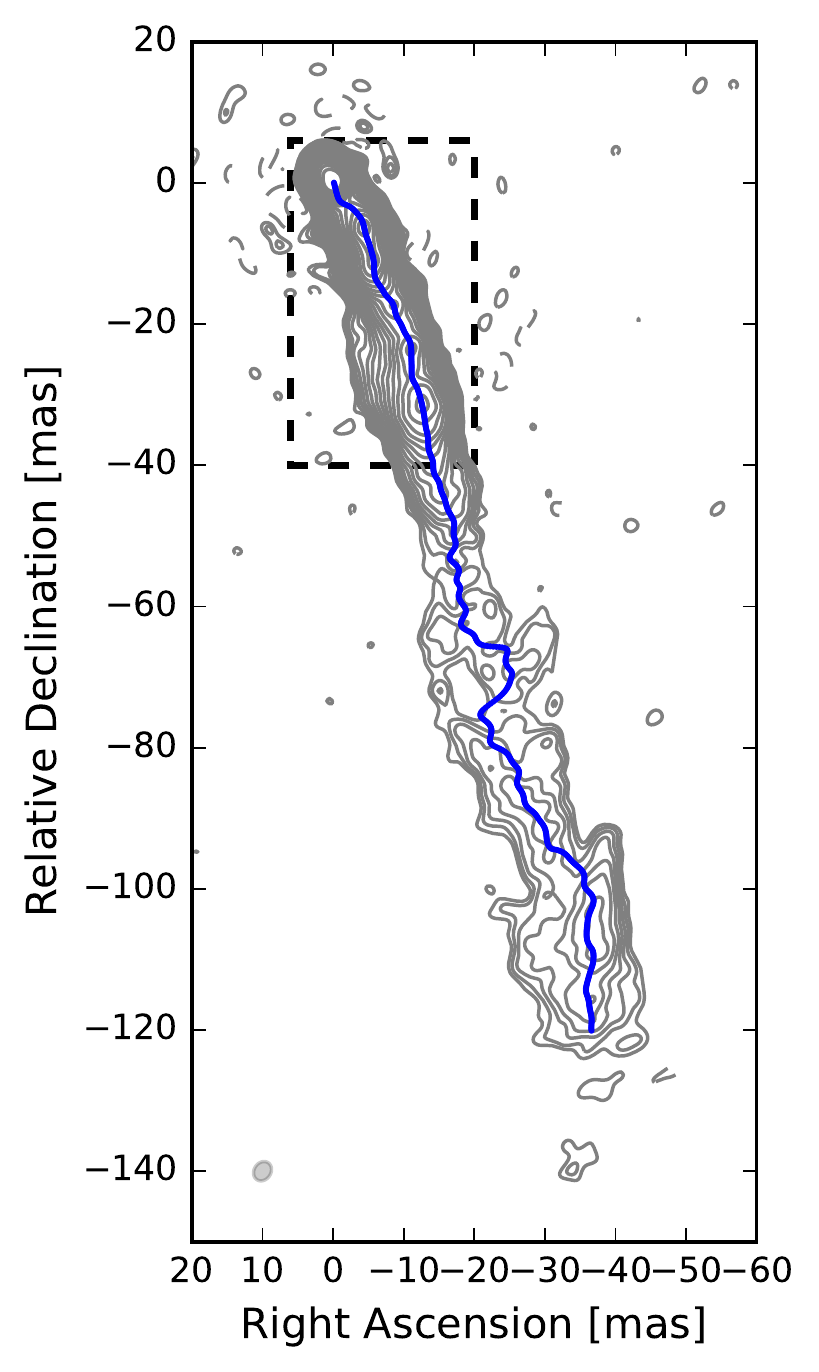}
  \caption{Ground VLBI image of S5\,0836$+$710 at 1.6 GHz. Contours are
    drawn at ($-$1, 1, $\sqrt{2}$, 2, etc. ) times 0.32~mJy/beam. Image
    parameters are listed in Table~\ref{tb:mapspar}. The dashed box
    marks the area covered by the respective {\em RadioAstron} image shown in
    Fig.~\ref{fg:0836-16-sridge}. The curved blue line denotes the jet
    ridge line derived and discussed in Sect.~\ref{sc:ridgeLineOscillations}.  }
\label{fg:0836-2-gridge} 
\end{figure}

\begin{table}[b!]
\caption{Parameters of the total intensity images.}
\label{tb:mapspar}
\begin{center}
\begin{tabular}{r|crrcr}\hline\hline
$\nu$ & $S_\mathrm{tot}$ & \mc{$S_\mathrm{peak}$} & \mc{$S_\mathrm{neg}$} & $\sigma_\mathrm{rms}$ & \mc{Beam} \\ \hline
    & \multicolumn{5}{c}{{\em RadioAstron} images} \\
1.6 & 3316  & 200    & $-$5.4  & 1.30 & 1.200, 0.210, $-$37 \\ 
5   & 3070  & 298    & $-$15.0 & 1.50 & 0.146, 0.056, $-$54 \\ 
22  & 1614  & 77     & $-$7.9  & 1.03 & 0.035, 0.016,\;\;\;77 \\  \hline
    & \multicolumn{5}{c}{Ground array images} \\  
1.6 & 3418  & 1160   & $-$0.8  & 0.15 & 2.880, 2.370, $-$34  \\
5   & 3373  & 1700   & $-$2.3  & 0.50 & 1.290, 0.975, $-$22\\
15  & 2429  & 1330   & $-$1.2  & 0.20 & 0.776, 0.567,\;\;\;\;\,9 \\
22  & 1658  & 603    & $-$1.6  & 0.30 & 0.388, 0.282, \;\,$-$6 \\
43  & 1171  & 538    & $-$3.4  & 0.60 & 0.349, 0.214, \;\,$-$3\\ \hline
\end{tabular}
\end{center}
{\small {\bf Column designation:}~$\nu$\,[GHz]: frequency. $S_\mathrm{tot}$\,[mJy]: total flux density. $S_\mathrm{peak}$\,[mJy/beam]:  peak flux density. $S_\mathrm{neg}$ [mJy/beam]: maximum negative flux density in the image. $\sigma_\mathrm{rms}$\,[mJy/beam]: rms noise in the image. Beam: major axis, minor axis, and position angle of major axis [mas,\,mas,\,$^{\circ}$] of the restoring beam.}
\end{table}

\begin{figure}[ht!]
  \centering
  \includegraphics[width=0.48\textwidth]{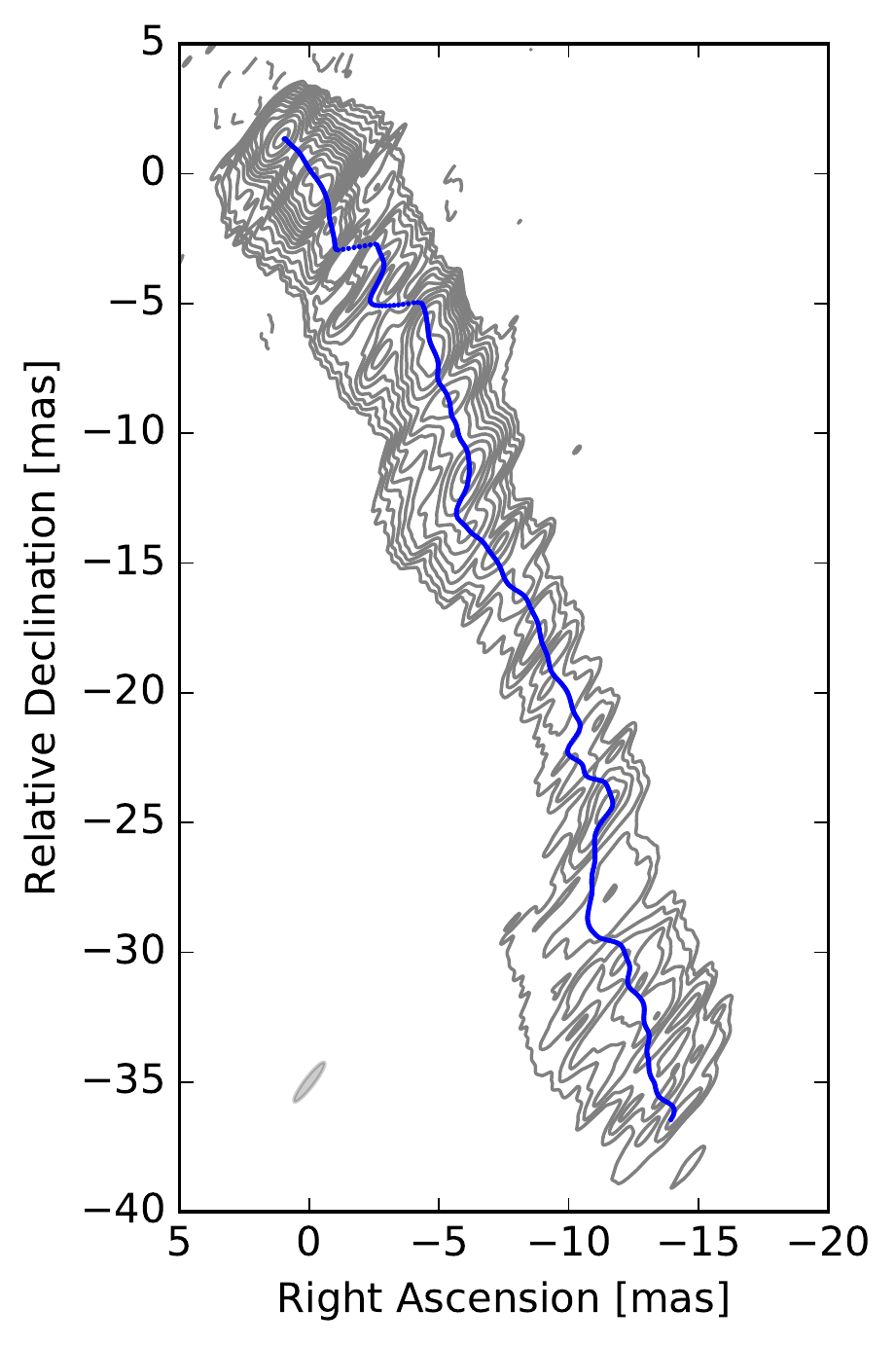}
  \caption{{\em RadioAstron} image of S5\,0836$+$710 at 1.6 GHz. Contours are drawn at ($-$1, 1, $\sqrt{2}$, 2, etc. ) times 10.0~mJy/beam. Image parameters are listed in Table~\ref{tb:mapspar}. The curved blue line denotes the ridge line we derived that is discussed in Sect.~\ref{sc:ridgeLineOscillations}.
  }
\label{fg:0836-16-sridge} 
\end{figure}

\begin{figure*}[t!]
 \centerline{
  \includegraphics[width=0.4385\textwidth,angle=0]{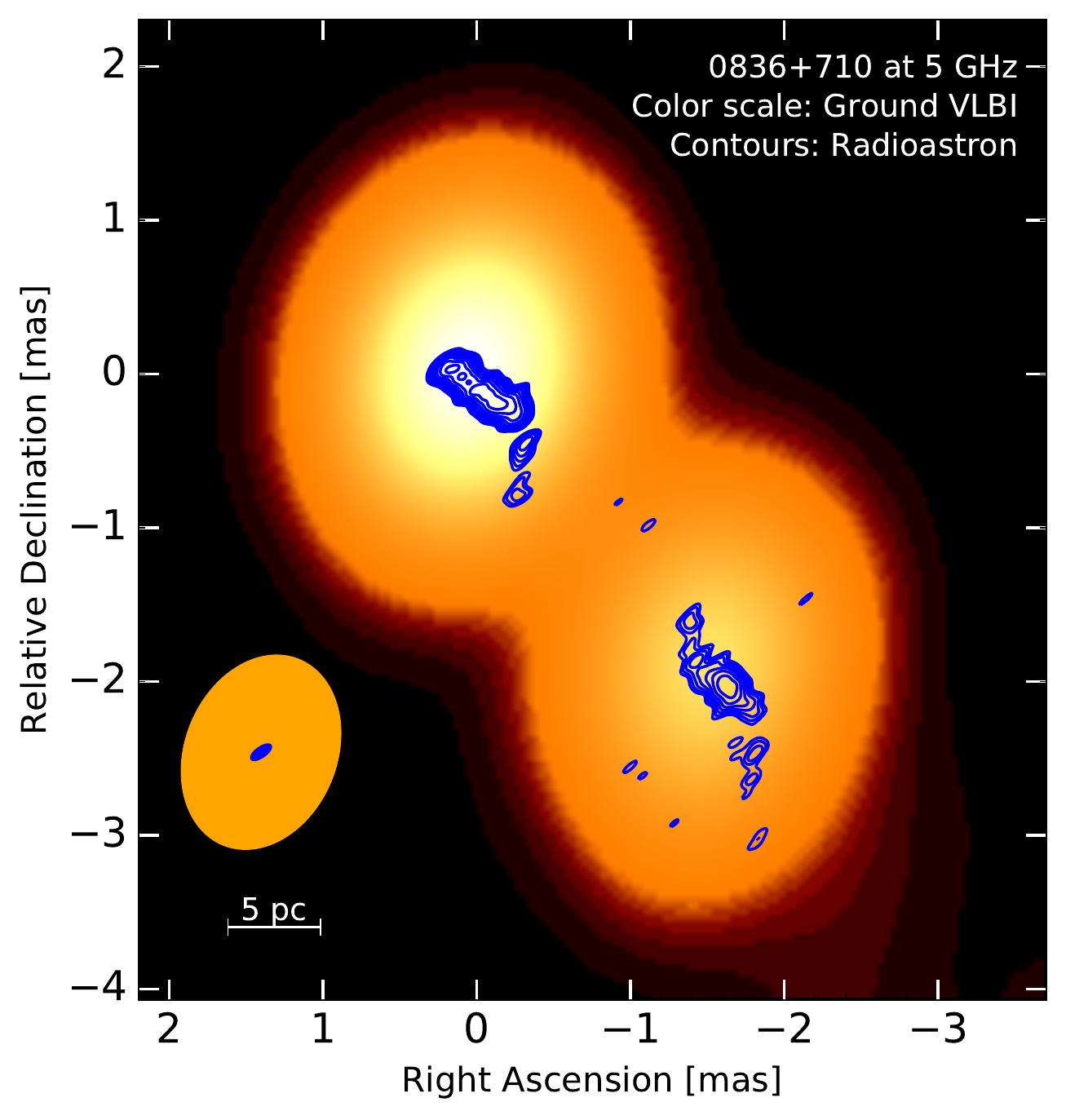}
  \includegraphics[width=0.5615\textwidth,angle=0]{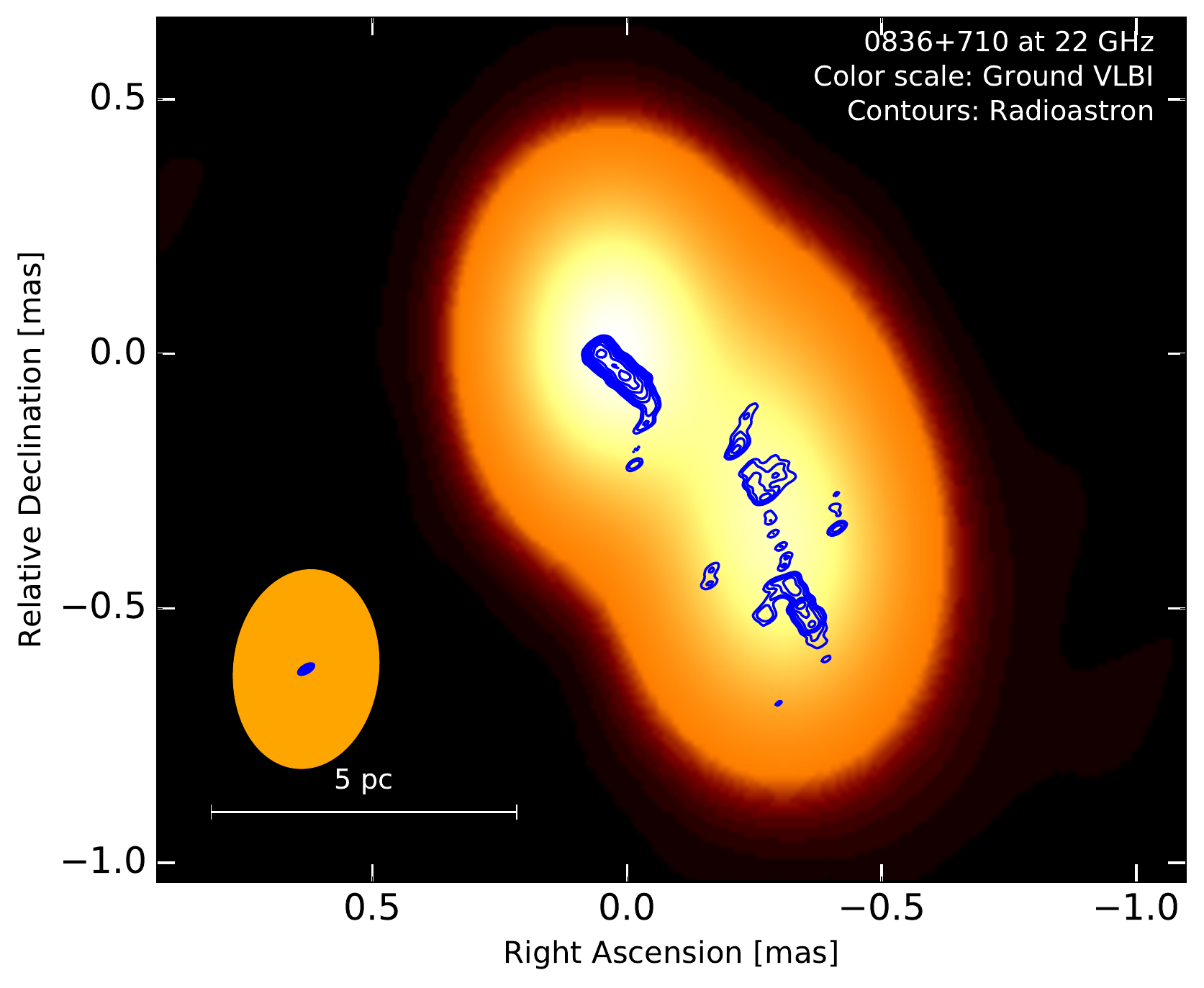}
}
  \caption{{\em RadioAstron} space VLBI images (contours) of
    S5\,0836$+$710 at 5\,GHz (left) and 22\,GHz (right) superimposed
    on images obtained at respective frequencies using only the ground-array data (colors). Ground-array images are made using
    natural weighting and space VLBI images are made using uniform data weighting.  The respective restoring beams are
    plotted in the lower left corner in orange (ground array) and blue
    (space VLBI).  Basic parameters of all images are
    listed in Table~\ref{tb:mapspar}.}
\label{fg:0836-5-22-ghz-maps} 
\end{figure*}

Figures~\ref{fg:0836-2-gridge}--\ref{fg:0836-5-22-ghz-maps} show the
resulting ground-array and {\em RadioAstron} images obtained from the
self-calibrated data at 1.6, 5, and 22\,GHz. The respective
ground array images at 15 and 43\,GHz are shown in
Figs.~\ref{fg:0836-15-gridge} and \ref{fg:0836-43-gridge} of
Appendix~\ref{ap:images}. The parameters of all of the 
individual images are listed in Table~\ref{tb:mapspar}.

\begin{figure*}[ht!]
\centerline{
\includegraphics[height=0.3\textwidth,angle=0,trim=25mm 0mm 25mm 3mm,clip=true]{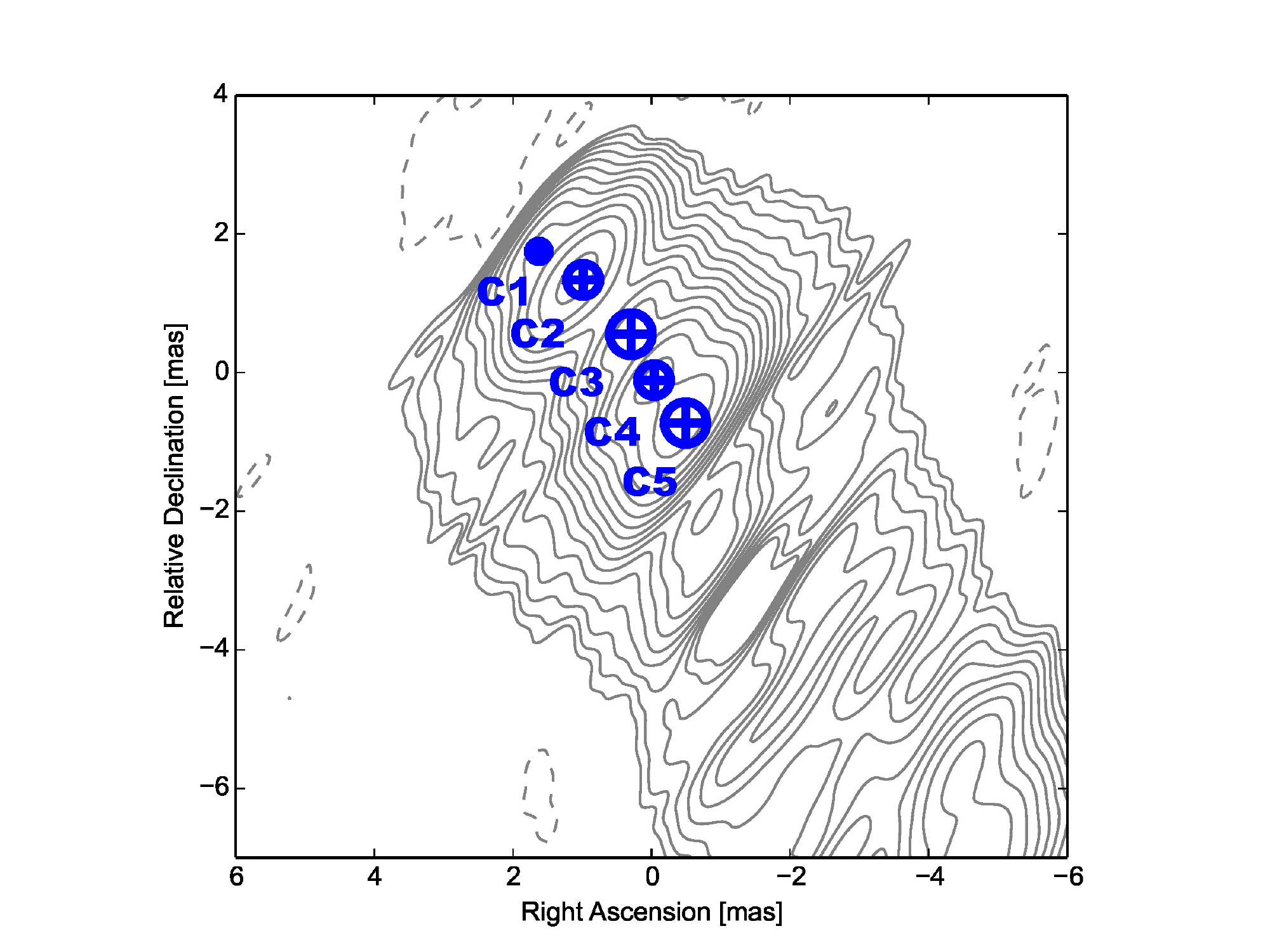}
\includegraphics[height=0.3\textwidth,angle=0,trim=30mm 0mm 25mm 3mm,clip=true]{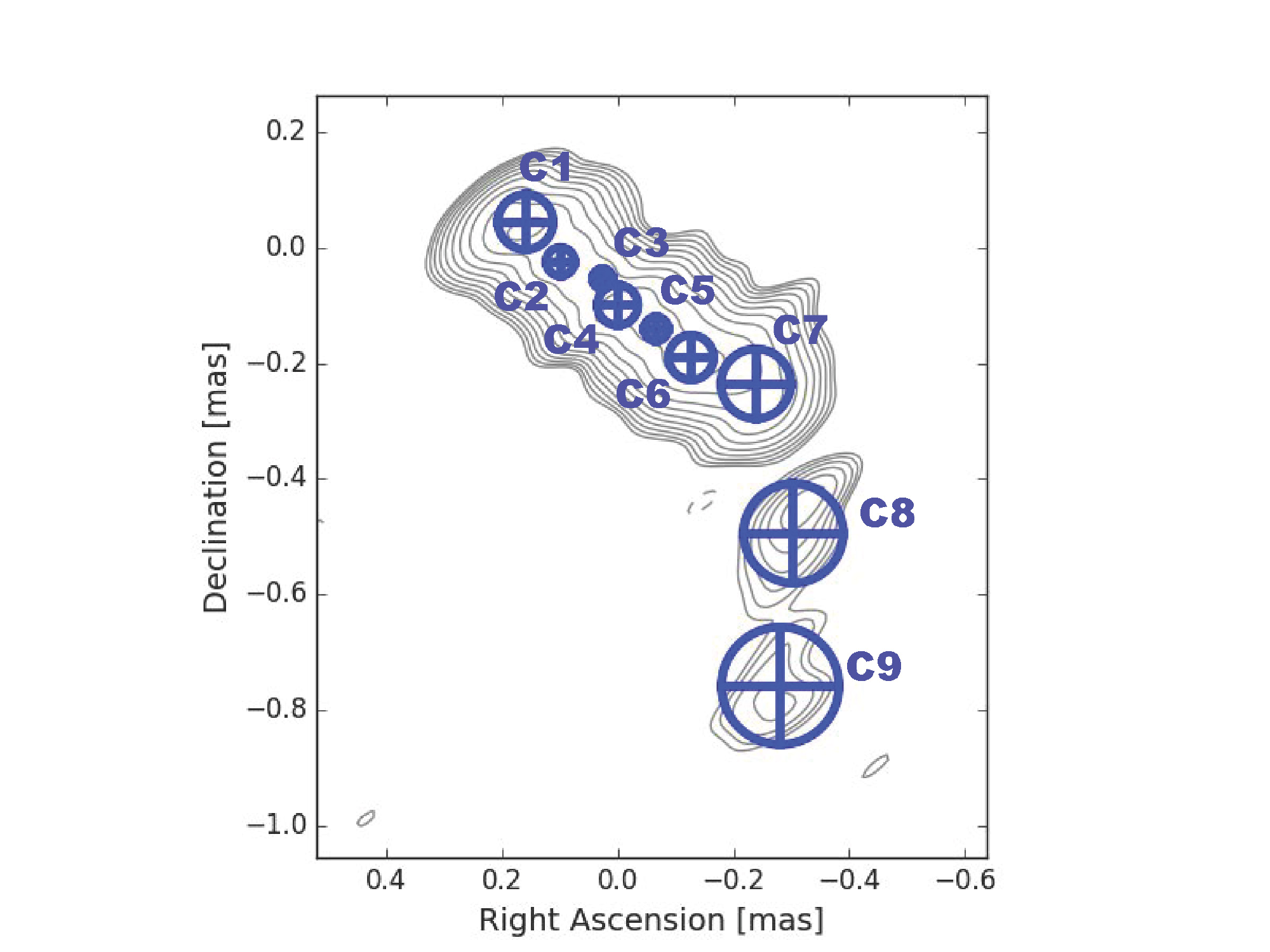}
\includegraphics[height=0.3\textwidth,angle=0,trim=30mm 0mm 35mm 3mm,clip=true]{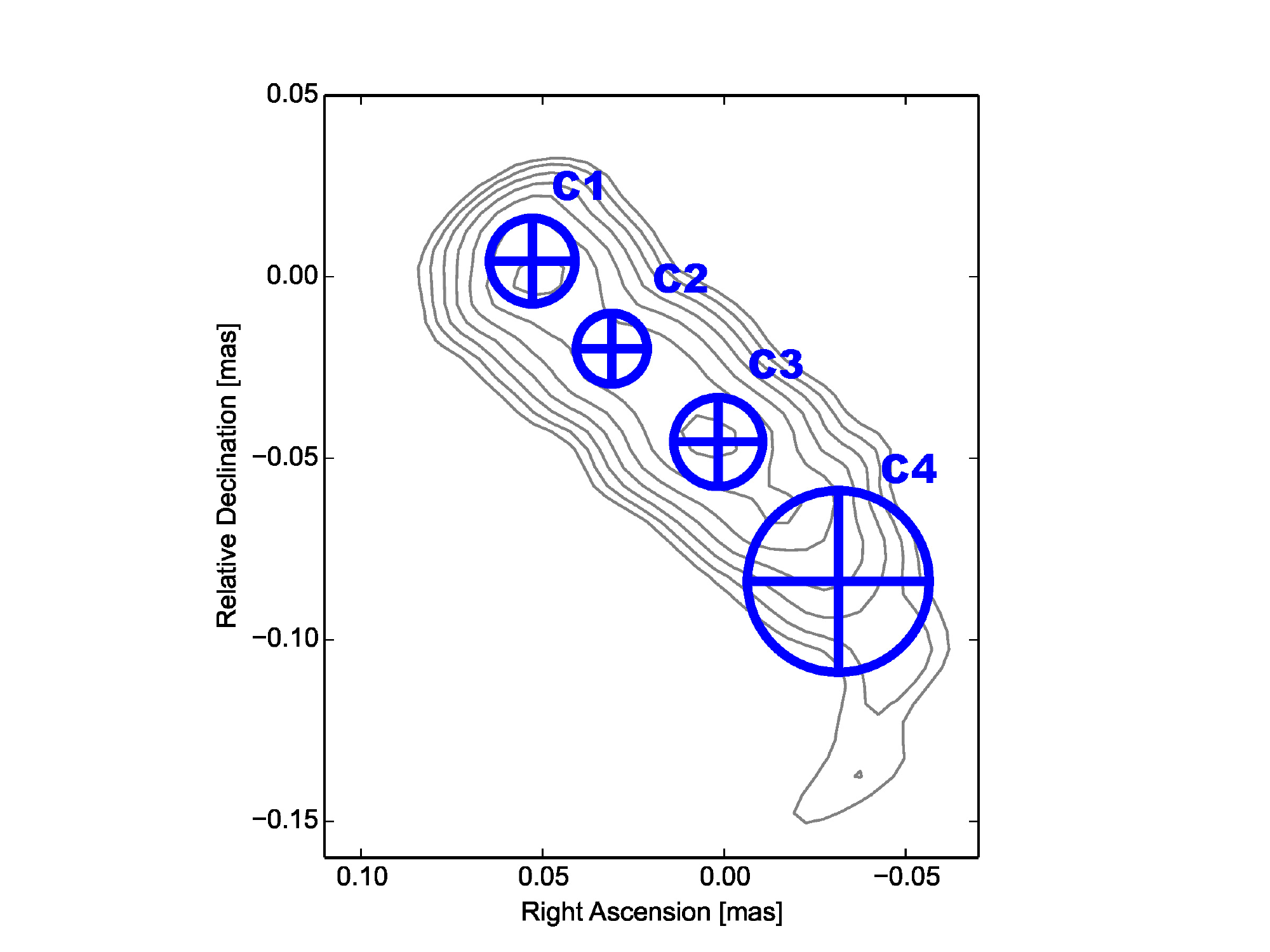}
}
\caption{Gaussian components used to represent the inner jet
  structure observed in the {\em RadioAstron} images of S5\,0836$+$710
  at 1.6\,GHz (left), 5\,GHz (center), and 22\,GHz (right). The
  derived component parameters are listed in Table~\ref{tb:modelfit}.}
\label{fg:0836-modelfit} 
\end{figure*}

The 1.6\,GHz images from the ground-array
(Fig.~\ref{fg:0836-2-gridge}) and space VLBI
(Fig.~\ref{fg:0836-16-sridge}) data show a remarkable wealth and
complexity of the structure, with the jet extending well beyond
40\,mas even in the full-resolution {\em RadioAstron} image.
The comparison of the space- and ground-VLBI images at 5 and 22\,GHz in
Fig.~\ref{fg:0836-5-22-ghz-maps} further demonstrates the remarkable
improvement in the angular resolution provided by {\em RadioAstron}. The
structure traced by the space baselines is located at the brightest
and most central part of the jet.  The jet seen at 1.6\,GHz shows
wiggles that could be part of a relativistic spiral with effects from
Kelvin-Helmholtz instabilities \citep[as suggested previously
in][]{lobanov+1998,perucho+2012a,perucho+2012b}.

The {\em RadioAstron} image at 5\,GHz shows that the jet is
substantially bent at small scales. Two clear bends are observed in
the two brightest regions. A Kelvin-Helmholtz instability developing in
the jet may explain this type of structure. The 22\,GHz {\em
  RadioAstron} image of S5\,0836$+$710 yields the highest resolution
image ever obtained for this source, reaching down to $\sim 15\,
\mu$as. This corresponds to a linear scale of $\sim 0.13$\,pc, or
$\sim 650$ gravitational radii for a black hole mass of $2 \cdot 10^9
M_\odot$ \citep{tavecchio+2000}.  Owing to the larger errors of the
visibility phase and amplitude as well as poor {\em uv} coverage on
the space baselines, the full-resolution 22\,GHz image can only trace
the brightest regions of the jet flow, with their morphology
suggesting a limb-brightened structure, which can be caused by
strongly asymmetric emission within the jet, or a
double-helical pattern, similar to that detected in 3C273 with VSOP
observations \citep{lobanov+2001}.

\begin{table*}[ht!]
\caption{Modelfit decomposition of the compact structure in S5\,0836$+$710 for space-VLBI observations.}
\label{tb:modelfit}
\begin{center}
\begin{tabular}{lcrrrrrr}\hline\hline
$\nu$ &Comp.& \mc{$S_\mathrm{tot}$} & \mc{$r$}   & \mc{$\phi$}       & \mc{$\theta$}          & \mc{$\theta_\mathrm{min}$} & \mc{$T_\mathrm{b}$}  \\
\,[GHz] &    & \mc{[mJy]}   & \mc{[mas]}        & \mc{[$^{\circ}$]} & \mc{[$\mu$as]} & \mc{[$\mu$as]} & \mc{[K]}  \\ \hline
1.6   & C1 &$ 70 \pm 15$ &$  2.39 \pm\;\, 0.07$ &$     43 \pm\;\, 2$ &$ 150 \pm\;\, 70$ & 107 &$1.4\times 10^{12}$ \\
      & C2 &$260 \pm 40$ &$  1.66 \pm\;\, 0.06$ &$     36 \pm\;\, 2$ &$ 240 \pm\;\, 60$ &  92 &$2.0\times 10^{12}$ \\
      & C3 &$170 \pm 50$ &$  0.63 \pm\;\, 0.12$ &$     28 \pm 11$ &$ 310 \pm 120$       & 124 &$7.9\times 10^{11}$ \\ 
      & C4 &$380 \pm 40$ &$  0.11 \pm\;\, 0.04$ &$   -160 \pm 20$ &$ 230 \pm\;\, 40$    &  77 &$3.3\times 10^{12}$ \\ 
      & C5 &$640 \pm 80$ &$  0.88 \pm\;\, 0.05$ &$   -146 \pm\;\, 3$ &$ 310 \pm\;\, 50$ &  84 &$3.0\times 10^{12}$ \\ \hline
5     & C1 &$350 \pm 10$ &$  0.16 \pm\;\, 0.02$ &$     74 \pm\;\, 4$ &$  50 \pm\;\, 20$ &  11 &$7.3\times 10^{12}$ \\ 
      & C2 &$ 90 \pm 10$ &$  0.10 \pm\;\, 0.02$ &$    104 \pm 13$ &$  25 \pm\;\, 18$    &  21 &$7.6\times 10^{12}$ \\ 
      & C3 &$310 \pm 15$ &$  0.10 \pm\;\, 0.03$ &$   -180 \pm\;\, 8$ &$  35 \pm\;\, 18$ &  14 &$1.3\times 10^{13}$ \\ 
      & C4 &$160 \pm 20$ &$  0.16 \pm\;\, 0.03$ &$   -160 \pm\;\, 3$ &$  22 \pm\;\, 18$ &  22 &$1.7\times 10^{13}$ \\ 
      & C5 &$290 \pm 30$ &$  0.23 \pm\;\, 0.02$ &$   -150 \pm\;\, 2$ &$  39 \pm\;\, 17$ &  20 &$1.0\times 10^{13}$ \\ 
      & C6 &$320 \pm 60$ &$  0.34 \pm\;\, 0.02$ &$   -140 \pm\;\, 3$ &$  60 \pm\;\, 20$ &  27 &$4.7\times 10^{12}$ \\ 
      & C7 &$150 \pm 60$ &$  0.06 \pm\;\, 0.02$ &$   -150 \pm 34$ &$  20 \pm\;\, 18$    &  38 &$>5.6\times 10^{12}$ \\ 
      & C8 &$115 \pm\;\, 8$ &$  0.58 \pm\;\, 0.08$ &$   -150 \pm 15$ &$  86 \pm\;\, 17$ &  17 &$8.2\times 10^{11}$ \\ 
      & C9 &$  6 \pm\;\, 1$ &$  0.81 \pm\;\, 0.08$ &$   -160 \pm\;\, 8$ &$ 100 \pm\;\, 20$ &  25 &$3.1\times 10^{10}$ \\ \hline
22    & C1 &$140 \pm 40$ &$ 0.053 \pm 0.007$ &$     85 \pm\;\, 8$ &$  12 \pm\;\,\;\, 7$ &   9 &$2.4\times 10^{12}$ \\
      & C2 &$110 \pm 30$ &$ 0.037 \pm 0.008$ &$    123 \pm 12$ &$  10 \pm\;\,\;\, 8$    &   9 &$2.7\times 10^{12}$ \\ 
      & C3 &$220 \pm 50$ &$ 0.046 \pm 0.005$ &$    178 \pm\;\, 7$ &$  12 \pm\;\,\;\, 5$ &   8 &$3.8\times 10^{12}$ \\ 
      & C4 &$120 \pm 40$ &$ 0.090 \pm 0.009$ &$   -159 \pm\;\, 6$ &$  25 \pm\;\,\;\, 9$ &  10 &$4.7\times 10^{11}$ \\ 
\hline
\end{tabular}
\end{center} {\small {\bf Notes:}~Gaussian model description: $S_\mathrm{tot}$
: total flux density. Component position in polar coordinates
($r,\,\phi$) with respect to the map center; component size, $\theta$.
Minimum resolvable size, $\theta_\mathrm{min}$, of the respective
component. Brightness temperature, $T_\mathrm{b}$, in the observer's
frame, derived for the parameters of the Gaussian fit.}
\end{table*}

\subsection{Brightness temperature}
\label{sc:tb}

We estimate the brightness temperature in the central regions of the
jet from two-dimensional Gaussian model fitting of the visibility
data, using the Levenberg-Marquardt algorithm implemented in {\em
  Difmap}. We compare the brightness temperatures calculated from the
Gaussian modelfits with visibility-based estimates \citep{lobanov2015}
obtained from the data at the longest ground-space baselines of the
observations. The Gaussian modelfits of the central regions and the
corresponding estimates of brightness temperature are shown in
Fig.~\ref{fg:0836-modelfit} and listed in Table~\ref{tb:modelfit}.

At 1.6\,GHz, a region of about 2.5\,mas in extent was modeled. At the
two higher frequencies, only the innermost part of the jet was
modeled, limited to the structure before the first gap of emission as
seen in Fig.~\ref{fg:0836-5-22-ghz-maps}. The model fitting was
performed in \textit{Difmap}, starting by removing the clean
components in the region of interest, and replacing them with a set of
fitted Gaussian components representing the respective structure.

At 1.6\,GHz (left panel of Fig.~\ref{fg:0836-modelfit}), five
components can describe the central region. For the innermost jet
structure observed at 5\,GHz and 22\,GHz (central and right panels of
Fig.~\ref{fg:0836-modelfit}), nine and four Gaussian components,
respectively, are needed to describe the most central part.  At
1.6\,GHz, the entire inner jet has a brightness temperature of
$\approx 10^{12}$\,K, with a maximum value of $3.3 \times
10^{12}$\,K estimated for the component C4 located downstream from the
apparent jet origin (see Table~\ref{tb:modelfit}). At 5\,GHz and
22\,GHz, the respective maximum brightness temperatures are also
achieved in components downstream in the jet, with $1.7 \times
10^{13}$\,K estimated for C4 at 5\,GHz and $3.8 \times 10^{12}$\,K for
C3 at 22\,GHz. This result underlines the complex structure of the
inner jet, with opacity \citep[see][]{lobanov1998,gomez+2016}, jet
acceleration \citep{lee+2016}, or fine-scale substructure \citep[e.g., resulting from recollimation
shocks; see][]{fromm+2013,lobanov+2015,gomez+2016} potentially responsible
for increasing the brightness downstream from the jet base.  In the
reference frame of S5\,0836$+$710, the respective brightness temperatures
are higher by a factor of $1+z$, hence all of them exceed
$10^{13}$\,K and require Doppler factors, $\delta_\mathrm{IC}
=20$--100 in order to reconcile them with the inverse Compton limit of
$\approx 5\times 10^{12}$\,K. Only the lower limit of the estimated
$\delta_\mathrm{IC}$ range agrees with the kinematic Doppler factor
$\delta_\mathrm{kin}\approx 17$ resulting from the existing measurements
of the jet speed and viewing angle \citep{otterbein+1998,lister+2013}.

\begin{table}[htb!]
\caption{Visibility-based estimates of the brightness temperature}
\label{tb:tbmin}
\begin{center}
\begin{tabular}{ccccc}\hline\hline
Band & $T_\mathrm{b,min}$  & $B$          & $\langle T_\mathrm{b,min}\rangle$ & $\langle B \rangle$ \\
     & [K]                 & [G$\lambda$] & [K]                       & [G$\lambda$] \\ \hline
  L  & $5.0\times 10^{13}$ & $0.64$       & $3.9\times 10^{12}$       & 0.66--0.73  \\
  C  & $3.4\times 10^{13}$ & $2.25$       & $2.9\times 10^{12}$       & 2.34--2.60  \\
  K  & $1.2\times 10^{13}$ & $11.2$       & $5.8\times 10^{12}$       & 10.4--11.5  \\ \hline
\end{tabular}
\end{center} {\small {\bf Notes:}~$T_\mathrm{b,min}$ : minimum brightness temperature obtained from the visibility data at a
given baseline length. $B$. $\langle T_\mathrm{b,min} \rangle$ : average
value of the minimum brightness temperature over the baselines in the
interval $\langle B \rangle$ corresponding to the longest 10\% of the
baselines in the respective data sets.}
\end{table}

Estimating the brightness temperature directly from the visibility
data (see Table~\ref{tb:tbmin}) yields even higher values, with the
minimum brightness temperature, $T_\mathrm{b,min}$ exceeding
$10^{13}$\,K at all three bands. When averaged over 10\,\% of the
longest baselines in each data sets, the values of $T_\mathrm{b,min}$
approach the modelfit estimates, but still remain higher. This
may be viewed as another indication that there is an ultracompact
structure in the jet that is not recovered by the Gaussian
modelfit. The visibility-based estimates require Doppler factors of up
to $\sim 300$ to avoid the inverse Compton catastrophe.  We
therefore conclude that the {\em RadioAstron} observations of S5\,0836$+$710 are
indicative of intrinsic violation of the Compton limit on the
brightness temperature.

Similar indications have also been seen in a number of other sources
observed with {\em RadioAstron}
\citep[see][]{lobanov+2015,gomez+2016,giovannini+2018}, and they might
imply that the innermost regions of the jet are either not in
equipartition or subject to more peculiar physical conditions
\citep{kellermann2002} such as a monoenergetic electron
energy distribution \citep{kirk+2006}. A more systematic study of the
overall {\em RadioAstron} measurements of the brightness temperature
should give better insights into this matter.

\subsection{Asymmetries of the jet structure}
\label{sc:asy}

The improved resolution of the {\em RadioAstron} images at 5\,GHz and
22\,GHz enables detailed studies of the innermost section of the flow,
on scales smaller than 1\,mas. The 22\,GHz image presented in
Fig.~\ref{fg:0836-5-22-ghz-maps} suggests that the jet is
transversally resolved, with a filamentary structure tracing only one
side of the flow. This apparent asymmetric transverse structure may
in part be due to the limited dynamic range (estimated to be about
75:1) of the images. It may also result from differential Doppler
boosting of the flow within the jet. Following \cite{Rybicki+1979}
\citep[see also][]{aloy2000,Lyutikov+2005,Clausen-Brown+2011}, the
bright side of the jet changes from top to bottom at a critical
viewing angle given by $\cos (\theta_\mathrm{r}) = \beta_\mathrm{j}$
for a helical magnetic field with its maximum asymmetry when the pitch
angle is $45^{\circ}$. In the case of S5\,0836+710, this corresponds
to $\theta_\mathrm{r} = 4^{\circ}$. Therefore, the top side of the jet
should be brighter, with the estimated jet viewing angle of
$3.2^{\circ}$ smaller than $\theta_\mathrm{r}$. This is indeed the
case, as observed in Fig.~\ref{fg:0836-5-22-ghz-maps}, implying a
helical field oriented counterclockwise
\citep[\textit{e.g.},][]{aloy2000}. Maximum asymmetry is obtained at
$\theta^{'} = \phi $, with $\theta^{'}$ the viewing angle in the jet
reference frame and $\phi$ the pitch angle in the fluid frame
\citep{aloy2000}. Fixing the viewing angle to $3^{\circ}$ and
$\beta_\mathrm{j}$ to that corresponding to Lorentz factor $\gamma =
12$ ($\beta_\mathrm{j}=0.99652$), we obtain $\phi \simeq 64^{\circ}$
for a maximum asymmetry. This pitch angle implies the presence of a
helical field with a slightly dominating toroidal
component. Nonetheless, the asymmetry in emission could also indicate
a physical asymmetry in the jet properties, revealing the presence of
distinct regions that vary across the jet channel with time. This has
been revealed by a number of VLBI observations that show components
that only partially fill the jet cross-section, and the jet is only
observed in its full width in stacked images over many epochs
\citep[\textit{e.g.},][]{lister+2013,pushkarev+2017,beuchert+2018}.

\begin{figure*}[htb!]
\includegraphics[width=0.4662\textwidth]{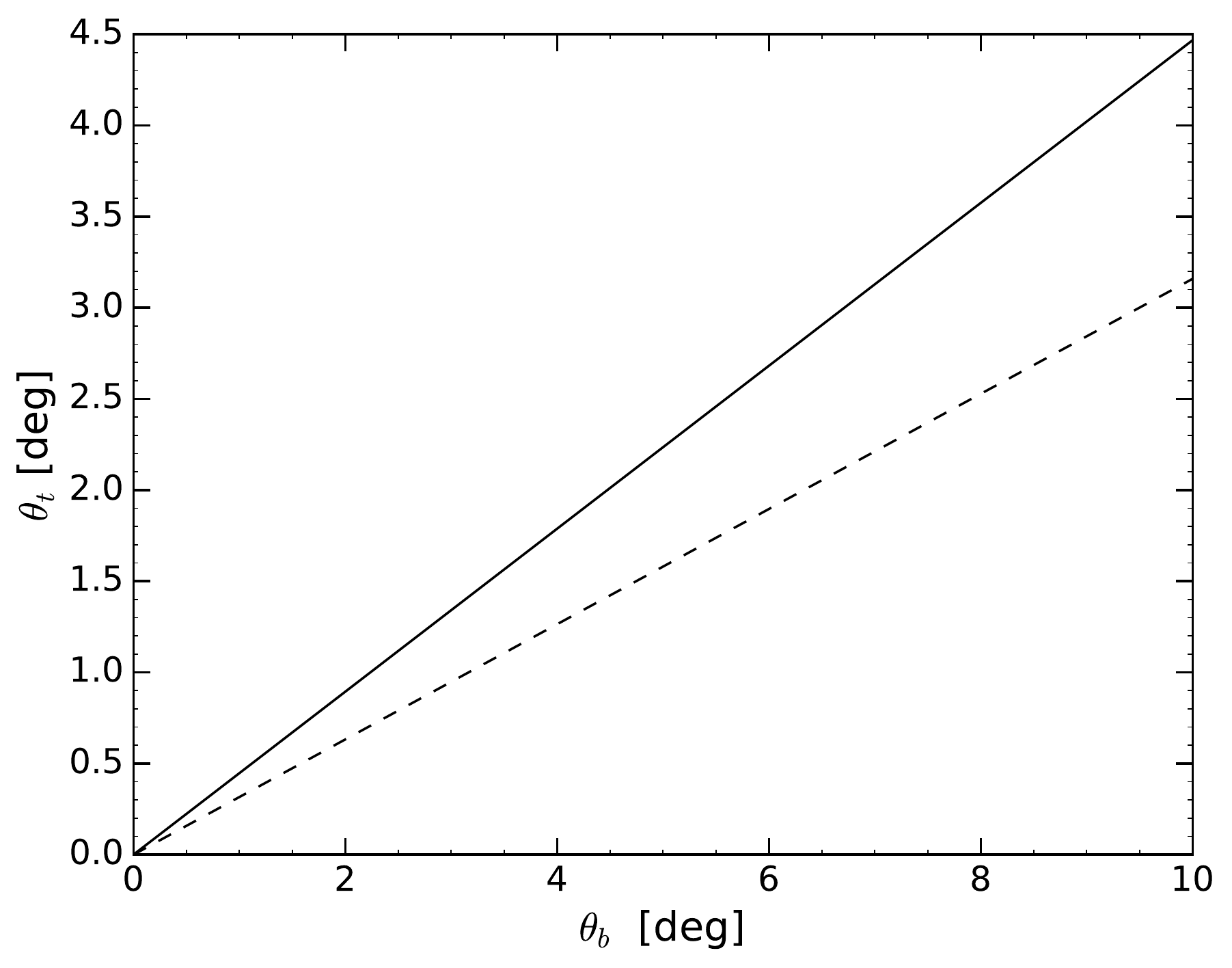}\,\includegraphics[width=0.4938\textwidth]{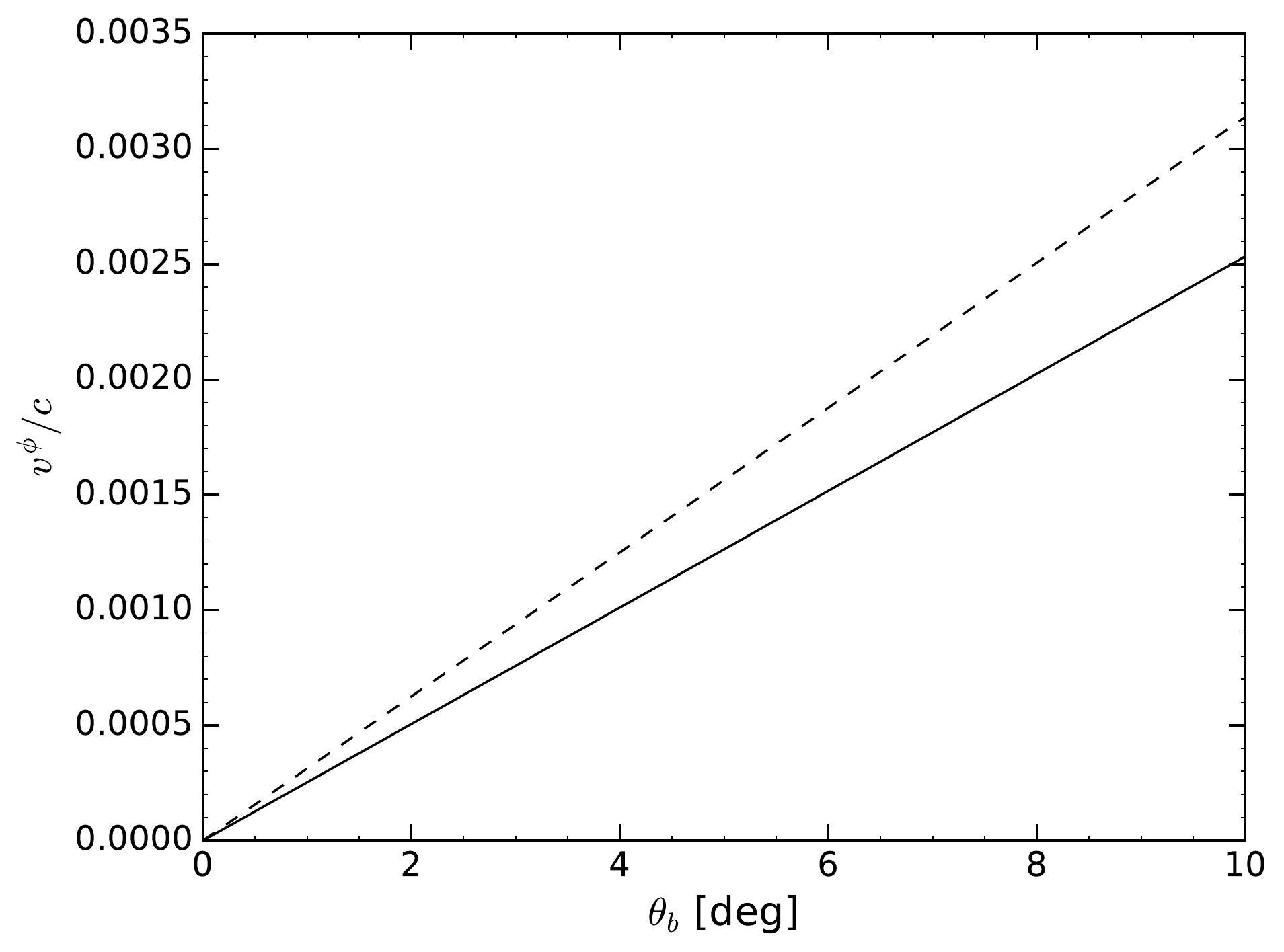}
\caption{Relations between the jet top- and bottom-edge viewing
  angles $\theta_\mathrm{t}$ and $\theta_\mathrm{b}$ (left panel) and
  the relative azimuthal velocity $v_{\phi}/c$ (right panel)
  required to explain a Doppler factor ratio of 5 (solid line) and 10
  (dashed line).}
 \label{fig:ang}
\end{figure*}

We can take the dynamic range of the 22\,GHz image as a measure of the
difference in brightness across the jet. Asymmetries in the transverse
structure of the jet emission could be caused by pressure differences
\citep{perucho+2007a,perucho+2012b}. However, this effect alone cannot
explain this difference in brightness if the perturbation is still
developing in the linear regime \citep{perucho+2005,perucho+2006}, and
the effect of the magnetic field orientation in the jet frame on the
synchrotron emissivity needs to be taken into account. Nonlinear
effects such as large instability amplitudes are not expected this
close to the jet base unless they respond to the development of
short-wavelength fast-growing modes
\citep[\textit{e.g.},][]{perucho+2007b}. However, in this case, we would expect a
persistent and symmetric emissivity. We thus exclude the option that
instability patterns generate this strong asymmetry in brightness.


We focus our discussion on the possibility that the asymmetry (of about the dynamic range of the observation) is produced by
differential Doppler boosting. The observed flux density is given by

\begin{equation}
S_\nu \propto D^{2+\alpha}\,B'^{1+\alpha}\,\sin(\theta'_\mathrm{B}),
\end{equation}
where $D$ is the Doppler factor, $B'$ and $\theta'_\mathrm{B}$ are the
intensity of the magnetic field and the angle between the magnetic
field and the line of sight in the fluid frame, respectively, and
$\alpha$ is the spectral index ($S_\nu \propto \nu^{-\alpha}$). In
this expression we ignore the dependence on the particle energy
distribution (in particular, on the Lorentz factor limits of this
distribution, $\gamma_{\mathrm{min}}$ and $\gamma_{\mathrm{max}}$,
which introduce dependencies on other parameters, such as the particle
number density). Recalling that the inner jet region has a flat
spectral index, $\alpha \simeq 0$ \citep[see][]{lobanov+2006}, and
assuming that the total magnetic field intensity (as given by both the
toroidal and poloidal components of the field) does not change much
with the toroidal angle (implicitly assuming axisymmetry), we obtain
the following expression for the brightness ratio between the top
brighter (at least by a factor of $\approx$75) part of the jet
(indicated with the superscript t) and the bottom part of the jet
(superscript b) in terms of the differential Doppler factors, $D$:

\begin{equation}   
    \frac{S_{\nu,\mathrm{t}}}{S_{\nu,\mathrm{b}}} \simeq \frac{D_\mathrm{t}^{2}
\sin(\theta'_\mathrm{B,t})}{D_\mathrm{b}^{2} \sin(\theta'_\mathrm{B,b})} \gtrsim 75\,,
\end{equation}
and in the following discussion, we assume this ratio to be $\sim 100$.

Taking into account that $\sin(\theta'_\mathrm{B}) \,= \, D \,\sin(\theta_\mathrm{B})$
(the unprimed value refers to the value in the observer's frame), we obtain

\begin{equation}  \label{eq:rat} 
    \frac{S_{\nu,\mathrm{t}}}{S_{\nu,\mathrm{b}}} \simeq \frac{D_\mathrm{t}^{3}
\sin(\theta_\mathrm{B,t})}{D_\mathrm{b}^{3} \sin(\theta_\mathrm{B,b})} \sim 100\,.
\end{equation}
The terms $D_\mathrm{t}$ and $D_\mathrm{b}$ can only be different if there is
a toroidal component of the velocity and henceforth the velocity
vector changes across the jet.

When we assume that the axial velocity component
$v_z \simeq c$ is constant
across the jet, the brightness asymmetry can be ascribed to the
top-to-bottom change of the angle
$\theta$ between the velocity vector and the line of sight, so
that from Eq.~\ref{eq:rat},

\begin{equation}   
    \frac{D_\mathrm{t}}{D_\mathrm{b}} \simeq
\frac{1-\cos(\theta_\mathrm{b})}{1-\cos(\theta_\mathrm{t})} \simeq 5
\left( \frac{\sin(\theta_\mathrm{B,b})}{\sin(\theta_\mathrm{B,t})} \right)^{1/3},
\end{equation}
             
\noindent which allows us to derive $\theta_\mathrm{t}$ in
terms of $\theta_\mathrm{b}$ if we approximate $ \left(
  \sin(\theta_\mathrm{B,b})/\sin(\theta_\mathrm{B,t}) \right)^{1/3}
\sim 1$ . The result is shown in the left panel of Fig.~\ref{fig:ang}
for the Doppler factor ratios of 5 (solid line) and 10 (dashed
line). In the plot, $\theta_\mathrm{b}$ spans from
$0^{\circ}$ to 10$^{\circ}$ in the observer's frame. The required
changes in the viewing angle can give estimates of the relative values
of the toroidal velocity component, $v_{\phi}$, as
$v_{\phi}/v_z\,=\,\tan(\psi)$, with $\psi$ defined as the deprojection
of the half-angle formed by the velocity vectors on both sides of the
jet (here we assumed a mean viewing angle of $3^\circ$ to
  deproject the half-angle between the velocity vectors). This ratio
is plotted in the right panel of Fig.~\ref{fig:ang} for the same
Doppler factor ratios. The plots show that low toroidal velocities
can result in sufficiently large changes in the Doppler factor of
different regions in the flow.

When we assume in contrast that $D_\mathrm{t}\simeq D_\mathrm{b}$, then the brightness
asymmetry has to be given by

 \begin{equation}   
    \frac{S_{\nu,\mathrm{t}}}{S_{\nu,\mathrm{b}}} \simeq \frac{
\sin(\theta'_\mathrm{B,t})}{\sin(\theta'_\mathrm{B,b})} \simeq 100\, .
\end{equation}

Taking into account that this ratio
$\sin(\theta'_\mathrm{B,t})/\sin(\theta'_\mathrm{B,b})\,\simeq\,\sin(\theta_\mathrm{B,t})/\sin(\theta_\mathrm{B,b})$
($D^\mathrm{t}\simeq D^\mathrm{b}$) and that $0 \leq
\sin(\theta_\mathrm{B}) \leq 1$, the fraction can only give a value
$\simeq 100$ if one or both values of $\theta_\mathrm{B}$ are very
close to $0^\circ$, where the sine can take values that differ by
several orders of magnitude. We find this coincidence that both angles
are aligned to such an accuracy less probable than the presence of low
toroidal velocities. Furthermore, a toroidal component of the velocity has
been shown to naturally arise in RMHD simulations of expanding jets
due to the Lorentz force \citep{marti+2016}. Finally, the required
azimuthal velocities are compatible with the axial velocity (Lorentz
factor) reported for this jet \citep{perucho+2012a} in terms of
causality.  We can conclude that a low toroidal velocity has to be
present in the jet flow to explain the brightness asymmetry. 
  Recent numerical GRMHD simularions of jet formation
  \citep{mckinney+2009,mckinney+2012} indicated that the rotational speed
  reaches up to the Keplerian speed, $v_\mathrm{c}(r_\mathrm{j})$, at
  a given jet radius, $r_\mathrm{j}$. For the reported black hole
  mass of $2\times 10^9\,\mathrm{M}_{\odot}$ \citep{tavecchio+2000}
  and a jet radius $r_\mathrm{j} \approx 0.05$\,mas ($\approx
  0.4$\,pc) estimated from the {\em RadioAstron} images, the
  respective $v_\mathrm{c} \approx 0.015$\,c is substantially higher
  than the $v_\phi$ estimates made above. This may result from a
  slowdown due to thermal mixing or shear, but likely uncertainties in
  all of the measured quantities involved in these calculations
  preclude us form making any firm conclusion on the matter. We plan
to run RMHD simulations to further investigate the potential physical
mechanisms that define the estimated rotation in this jet.

\section{Jet structure: Ridge line oscillations}
\label{sc:ridgeLineOscillations}

\subsection{Ridge line calculation}
\label{sc:ridgeLines}

Because of the limited dynamic range of the space VLBI images at
5\,GHz and 22\,GHz (in which the jet is transversally resolved), we
base our quantitative analysis of the jet flow largely on the jet
ridge line derived from the 1.6\,GHz images of S5\,0836+710. At the
scales sampled by these images, plasma instability is expected to play
an important role, inducing regular patterns into the flow. We study
these patterns by estimating the ridge line of the jet. We define the
jet ridge as the line that connects the peaks of one-dimensional
Gaussian profiles fitted to the profiles (slices) of the jet brightness drawn
orthogonally to the jet direction \citep[see][for similar previous
studies]{lobanov+1998,perucho+2012a}, with the step between individual
profiles set to be smaller than the beam size (a measure taken in
order to ensure continuity of the brightness profiles recovered from
adjacent slices).  Similarly to the earlier studies of S5\,0836+710, a
position angle of $-$162$^\circ$ was adopted for the jet direction.  To
obtain the ridge line, a code was developed using \texttt{Python}
 and the \texttt{iMinuit}
package\footnote{https://pypi.python.org/pypi/iminuit}. The fitting
algorithm implements continuity conditions for the parameters fitted
to adjacent slices, thus ensuring a robust and self-consistent
description of the evolution of the ridge line and flow width along
the jet. The first slice is taken across the jet so as to cross the
peak of brightness in the image at a given frequency.  We calculated
the uncertainties of the fitted parameters from the S/N in the image
using the method described in \cite{schinzel+2012}.

We present the ridge lines obtained from
the ground- and space-VLBI images of S5\,0836+710 at 1.6\,GHz in
Figs.~\ref{fg:0836-2-gridge}--\ref{fg:0836-16-sridge}. The ridge lines
obtained from the images at other frequencies are presented in
Appendix~\ref{ap:images}.

The complex ridge line structure suggests the presence of several
periodic patterns in the flow. In the following, we assume that these
oscillatory modes represent individual modes of Kelvin-Helmholtz (KH)
instability developing in the flow
\citep[see][]{lobanov+2001,perucho+2007b,perucho+2012a}.

\subsection{Modeling  the KH instability modes}
\label{sc:modelling}

\begin{figure*}[t!]
\includegraphics[width=0.5\textwidth]{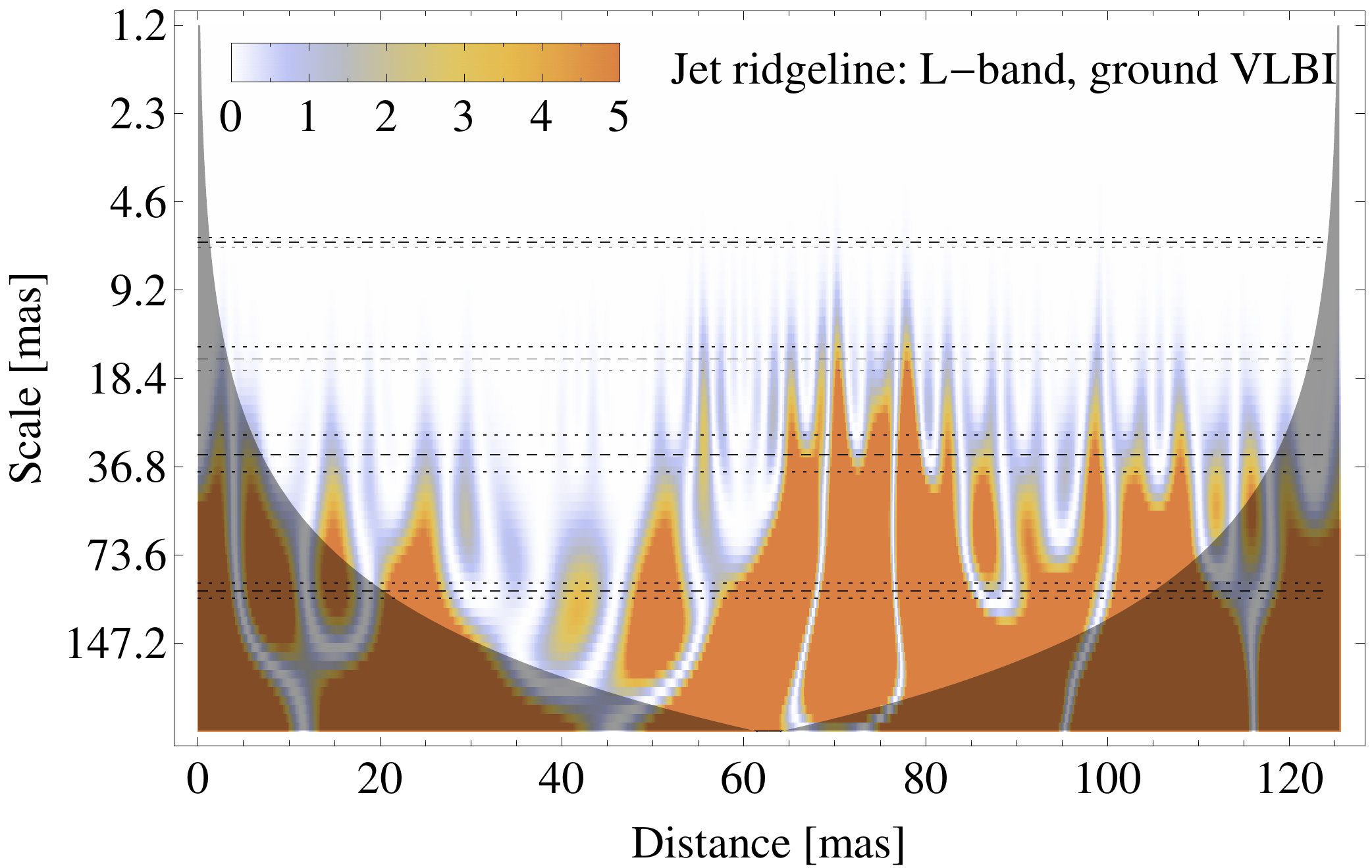}\,\includegraphics[width=0.5\textwidth]{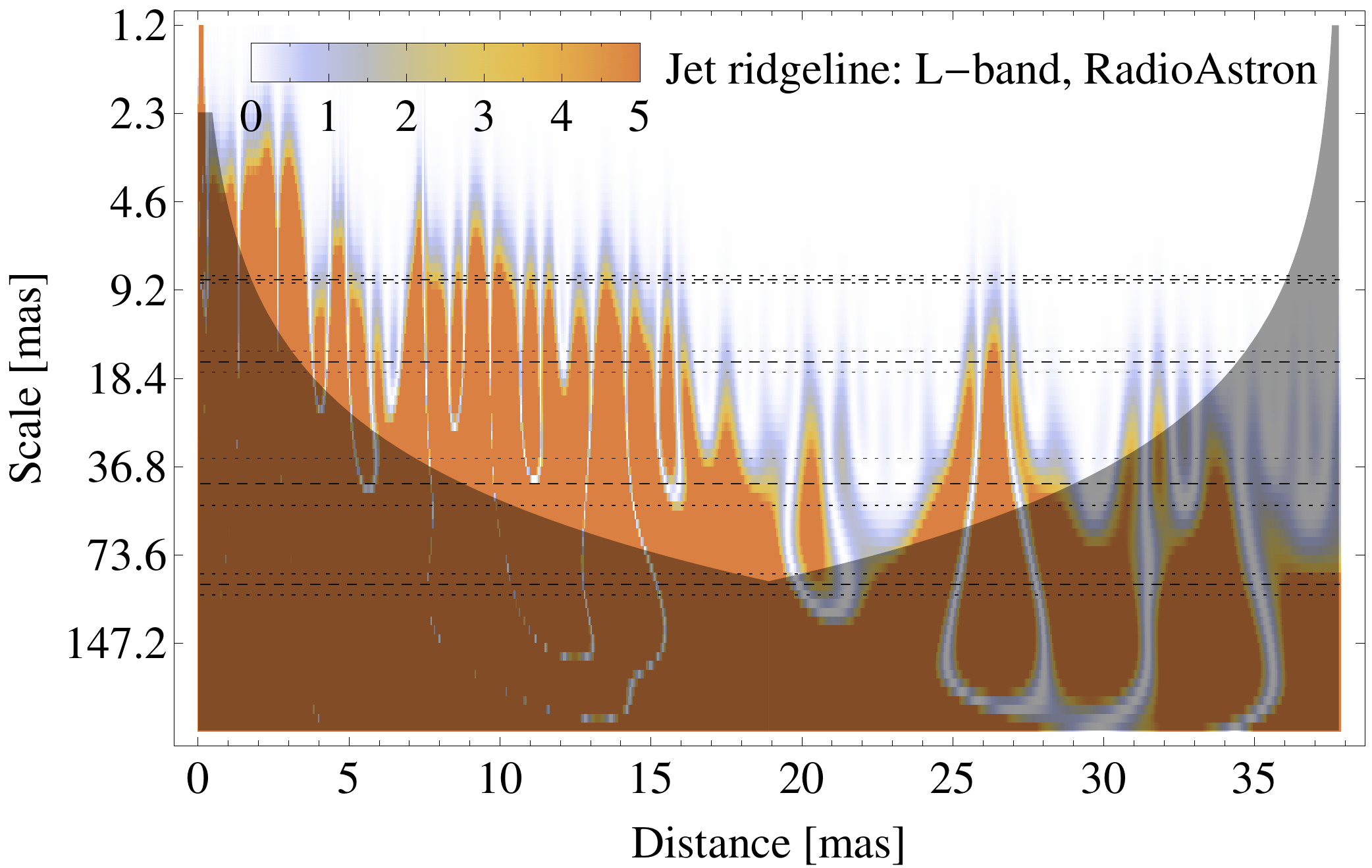}
\caption{Wavelet scalograms for the ridge lines of S5\,0836+710 at
  1.6\,GHz, obtained from the ground- (left) and space-VLBI (right)
  images. Color presents the wavelet amplitude scaled with the mean
  error, $\sigma_\mathrm{pos}$, of the ridge line position
  measurements. The color scale is saturated at
  $5\,\sigma_\mathrm{pos}$ level for better viewing. Undersampled
  regions of the parameter space are shaded. For both ridge lines, the
  scalogram is dominated by an oscillation with a long wavelength of
  $\sim 90$--$100$~mas, although it cannot be well sampled with the
  ridge line from the space-VLBI image. Presence of one or more
  oscillatory patterns with wavelengths of $\sim 25$--$45$~mas can
  also be suggested from both scalograms. A mode with the wavelength
  of $\sim 10$--$15$~mas can be inferred from the space-VLBI
  ridge line. For the purpose of comparison, each scalogram also shows
  the wavelengths (dashed lines) and respective uncertainties (dotted
  lines) of the oscillatory modes fitted to represent the ridge lines
  as described in Section~\ref{sc:modelling}.  For each of these
    wavelengths, the respective scalogram contains at least one region
    of statistically significant wavelet amplitudes, thus reaffirming
    the physical relevance of the fitted oscillatory modes.}
 \label{fig:wavelet}
\end{figure*}

We initially identifed the potential modes using a
wavelet scalogram calculated from the ridge line offsets measured in
the ground- and space-VLBI images at 1.6 GHz (Fig.~\ref{fig:wavelet}),
with the Marr function as the kernel for the wavelet transform. The
scalograms presented in Fig.~\ref{fig:wavelet} indicate
oscillatory modes at wavelengths of $\sim$\,90\,mas, $\sim$\,35\,mas,
$\sim$\,15\,mas, and $\sim$\,10\,mas. The wavelet scalograms also show
that the oscillatory patterns may have a slightly increasing
wavelength with distance, a typical behavior expected for a KH
mode propagating in an expanding flow \citep{hardee2000}. These
inferences are used as initial guesses for further modeling of the
ridge lines.

We used the ridge line measured in the 1.6\,GHz ground- and space-array images as the basis for our modeling. We first describe the
observed offset, $\Delta\,r(z)$, of the ridge line with a simplified
model by fitting it with several different oscillatory terms with
constant amplitudes and wavelengths:
\begin{equation}
\Delta\,r (z) = \sum_{i=1}^{N_\mathrm{mod}} a_i \sin(2\,\pi/\lambda_i + \psi_i)\,,
\end{equation}
where $a_i$ is the amplitude, $\psi_i$ the phase and $\lambda_i$ the
wavelength corresponding to the $i$-th mode. Four modes were needed to minimize the $\chi^2$ in both ridge lines. The fitting was performed in two ways:

\begin{enumerate}
\item Fitting the first mode and subtracting it from the original
  ridge line. Then, a second mode was fit and again subtracted. The
  process was iterated until the addition of a new mode did not
  provide any improvement to the fit.
\item Fitting all modes at the same time. This method  depends more strongly on the initial guess parameters. The first three modes
  were fit with the wavelengths observed in the scalogram as an
  initial guess.
\end{enumerate}

\begin{figure*}[t!]
  \includegraphics[width=0.48\textwidth]{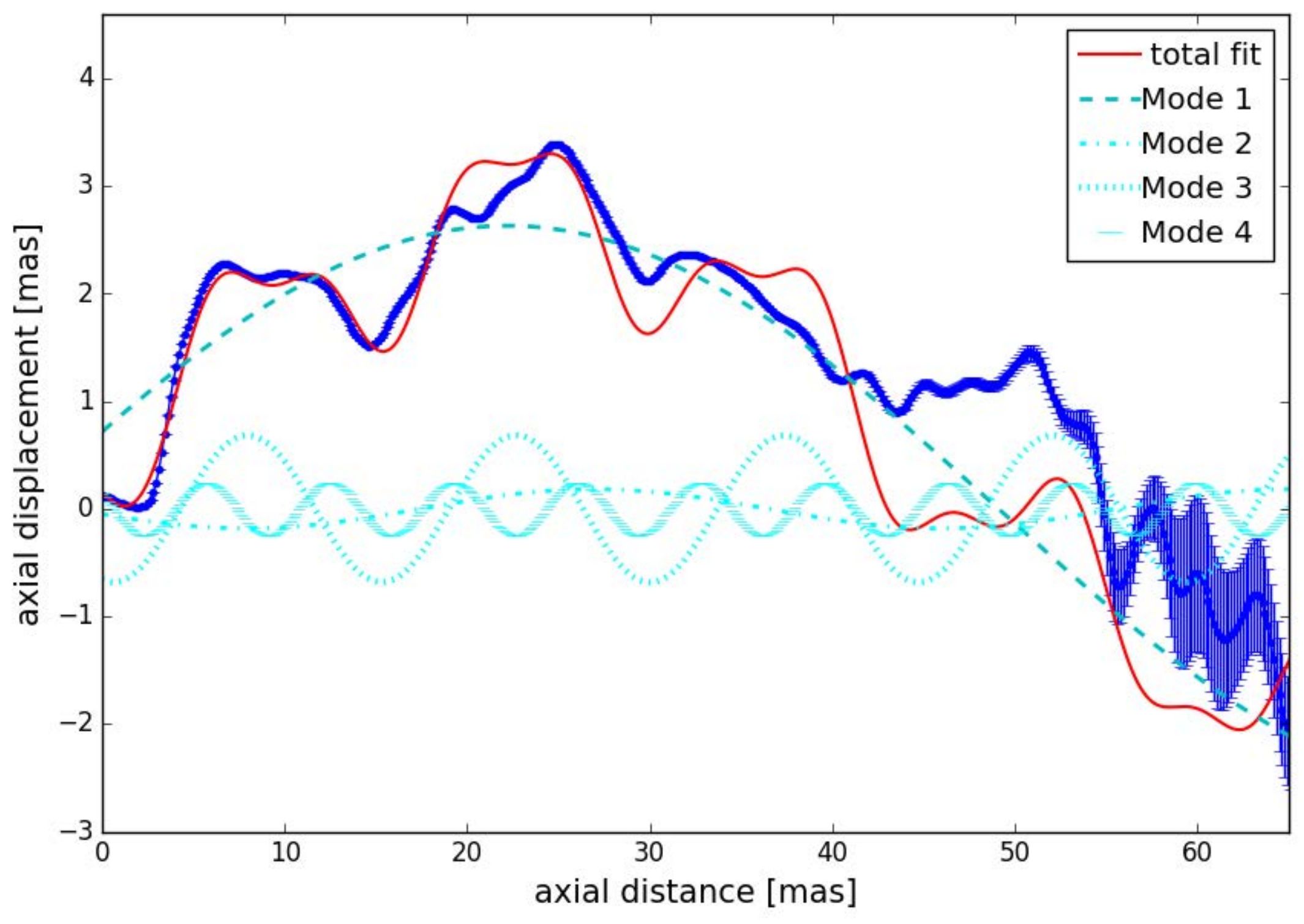}
  \includegraphics[width=0.48\textwidth]{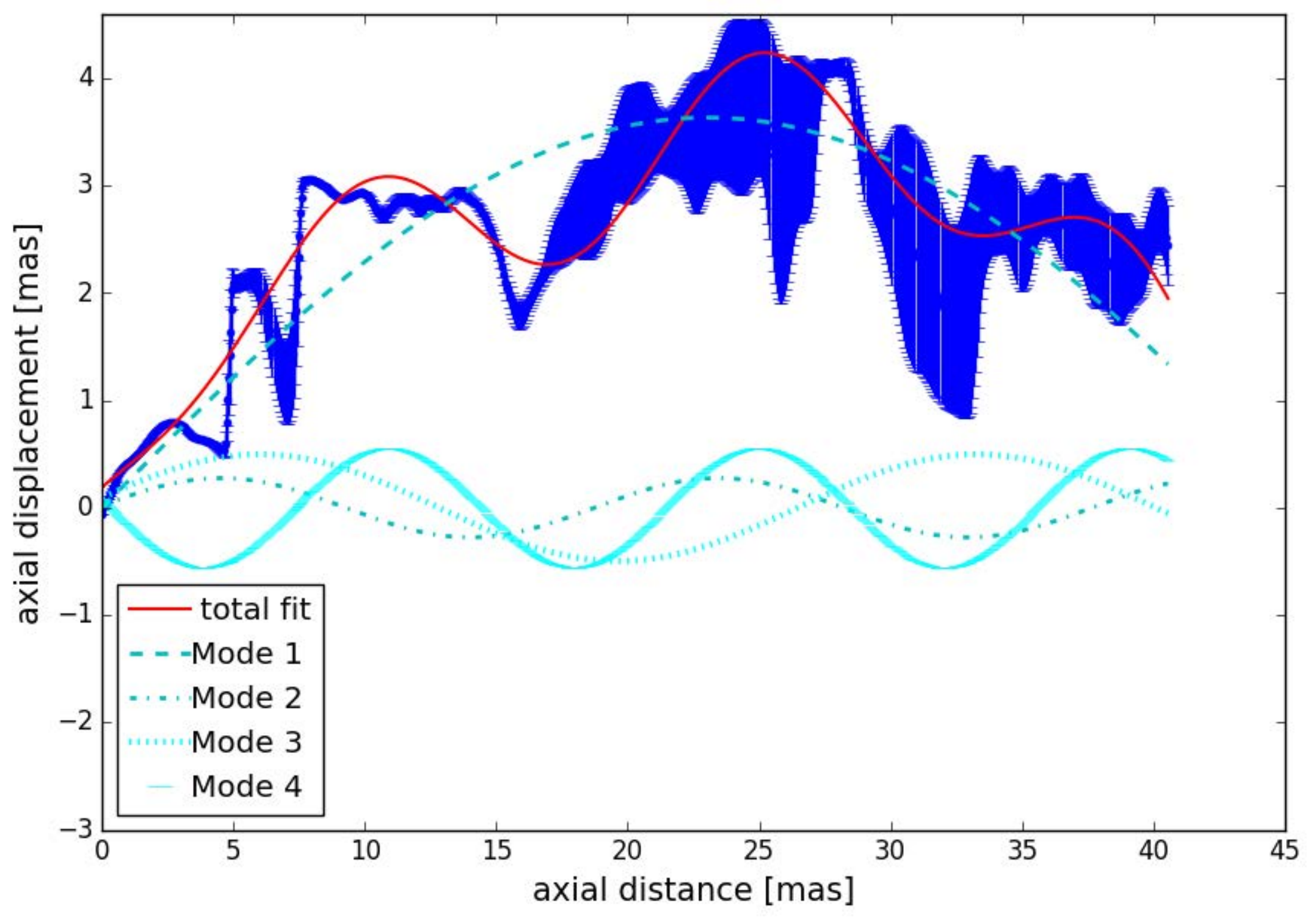}
\caption{Fits by oscillatory modes to the jet ridge lines in the ground- (left) 
and space- (right) VLBI images of S5\,0836+710 at 1.6\,GHz. In each plot, the 
red line is the total multi-mode fit and the blue lines represent contributions from the individual oscillatory modes as described in the legend.
  \label{fig:modesfit}}
\end{figure*}

For both methods, the best fit was chosen to be that which reduced the
$\chi^2$. The two methods yield similar results. The wavelengths of
the modes are listed in Table~\ref{tb:modes16} for the ground-image
ridge line and in Table~\ref{tb:modes16space} for the space-VLBI image
ridge line. Figure~\ref{fig:modesfit} shows the resulting fit for the
two ridge lines. The resulting wavelengths obtained
from two independent fits are similar in both cases. This corroborates
the robustness of our identification of the oscillatory modes and the
little role played by jet expansion on the wavelength, taking into
account the small jet opening angle
\citep[$<1^\circ$][]{perucho+2012a}.

\begin{table}[!tbhp]
\centering
\caption{Oscillatory mode representation of the ridge line in the ground-array L-band image}
{\begin{tabular}{c|rrr}
\hline \hline
Mode &  $\lambda$ [mas]& $a$ [mas] & $\psi$ [$^\circ$] \\
\hline
\hline
 1     & $102.0 \pm 6.0$ &  $2.90\pm 0.20$ & $15\pm \;\;8$ \\
 2     & $35.0 \pm 5.0$ &  $0.21\pm 0.03$ & $195\pm 50$ \\
 3     & $16.5 \pm 1.5$ &  $0.55\pm 0.16$ & $-105\pm 30$ \\
 4     & $6.6\pm 0.2$ &  $0.30\pm 0.05$ & $146\pm 11$ \\
\hline
\end{tabular}
}
\label{tb:modes16}
\end{table}

\begin{table}[!tbhp]
\centering
\caption{Oscillatory mode representation of the ridge line in the {\em RadioAstron} L-band image}
{\begin{tabular}{c|rrr}
\hline \hline
Mode &  $\lambda$ [mas]& $a$ [mas] & $\psi$ [$^\circ$] \\
\hline
\hline
 1     & $97.0\pm 4.0$ &  $3.70\pm 0.20$ & $0\pm 10$ \\
 2     & $44.0\pm 8.0$ &  $0.42\pm 0.04$ & $82\pm 14$ \\
 3     & $16.9\pm 1.4$ &  $0.50\pm 0.12$ & $-106\pm \;\;7$ \\
 4     &  $8.8\pm 0.1$ &  $0.23\pm 0.07$ & $54\pm \;\;6$ \\
\hline
\end{tabular}
}
\label{tb:modes16space}
\end{table}

When we compare these results with earlier works, it is interesting to note
that the wavelengths identified in the ridge line at 1.6\,GHz agree well with those that were obtained previously from the VSOP image
of the source \citep{lobanov+1998}, where oscillations with
wavelengths of $\simeq 100$, $\sim 8$, and $\sim 5$\,mas were
identified. The longest of these wavelengths is reasonably close to
the $\lambda = 97$ or $\lambda = 102$\,mas of the mode 1 in our
\emph{RadioAstron} data, and the two shorter wavelengths are close to
$\lambda = 6.6$\,mas from the ground-array image and $\lambda =
8.8$\,mas of the mode 4 from the space-VLBI image. It should be noted
that identification of short wavelengths in a ridge line measured in
different VLBI images can be affected by the differences in the image
noise and the {\em uv} coverage of the observations. It is therefore
more difficult to give a robust representation of short-wavelength
oscillations in a ridge line. We note that a $\simeq 80$\,mas mode was
also detected in VLBA images of S5\,0836+710 by \citet{perucho+2012a},
who suggested the presence of oscillations with shorter and possibly
variable wavelengths, and reported oscillations with a wavelength
between $\simeq 10$\,mas and $\simeq 20$\,mas at $r\leq40$\,mas
distance from the jet origin and growing to $\simeq 40$\,mas at
$r\geq40$\,mas.
Finally, \cite{perucho+2012a} also reported the detection of
short-wavelength oscillations, with wavelengths $\simeq 5$\,mas at
$r<15$\,mas increasing to $\simeq 7$--$8$\,mas at $r>20$\,mas. These can
be related to mode 4 in Tables~\ref{tb:modes16} and~\ref{tb:modes16space} (wavelengths of 6.6 and 8.8 mas).  It is
also relevant to stress that in our fits we did not include growth
lengths for the modes, which can also introduce small differences with
the data at the largest scales. Based on all these arguments, we
conclude that our results are consistent with previous studies of the
instability patterns detected in the jet of S5\,0836+710.

Regarding the possible identification of the instability modes,
because the first, longest-wavelength mode displaces the
ridge line, it should correspond to a helical mode. It could be
tentatively identified as an helical surface mode, $H_\mathrm{s}$.

A tentative identification of the other modes can be obtained using
the characteristic wavelength,
$\lambda^*=\lambda_i(n_i+2m_i+\frac{1}{2})$, where $\lambda^*$ is the
characteristic wavelength, $\lambda_i$ the observed wavelength and
$n_i$ and $m_i$ the two transverse wavenumbers of the mode. The
characteristic wavelength should have a similar value for the
different modes \citep{lobanov+2001}, assuming that all the observed
wavelengths correspond to modes that are excited at their maximum growth
rates. The second-longest wavelength should then be the first helical
body mode, $H_\mathrm{b1}$, whereas the third longest wavelength should be
either the second, $H_\mathrm{b2}$, or the third helical body mode,
$H_\mathrm{b3}$. We exclude the fourth mode because such small wavelengths are
difficult to determine and the results for ground and space arrays do
not reconcile, even when the errors are accounted for. For the case in which
the third mode is identified as the second-order helical body mode,
the characteristic wavelength is $119 \pm 12$\,mas, and when the third mode is identified as the third-order body mode, the
characteristic wavelength is $130 \pm 12$\,mas.


Following the approach of \citet{hardee2000} and references therein,
the basic physical parameters of the jet can be obtained from the
identification of the different modes, using the linear analysis of
the Kelvin-Helmholtz instability. With this approach, it is possible to
obtain the Mach number of the jet and the ratio of the jet to the ambient density
knowing the characteristic wavelength, the jet radius, jet viewing
angle, the jet apparent speed, and the apparent pattern speed. The jet
radius can be easily calculated.  It corresponds to the point where
the peak flux density of the jet is reduced to $1\%$ along the jet
direction \citep{wehrle+1992}.

In practice, to choose the slice where we measure the jet width, we
considered the slice with a peak flux density $1\%$ of the flux density
of the first slice, corresponding to the brightest point of the
jet. For this slice, we took the full width at half-maximum (FWHM) of the
corresponding Gaussian profile as a measure of the jet width. Then we
deconvolved it, using the following relation: $R_\mathrm{j} {\mathrm{
    [mas]}} = 0.5 \sqrt{\theta_{\mathrm{FWHM}}^2 - b^2}$, where
$\theta_{\mathrm{FWHM}}$ is the FWHM of the
corresponding slice and $b$ is the beam size. This yields a jet radius of
$2$~mas or $16$~pc. We note that in earlier works
\citep{perucho+2007b,perucho+2011} a substantially higher value of
17\,mas was estimated for the jet radius. This value corresponds to
the apparent jet radius measured directly from the image, while we
deconvolved this measurement with the FWHM of the resolving beam. The
large radius considered in earlier works led to the identification of
the longest observed wavelength with the first body mode (because the observed structures are relatively shortened when they are measured with
respect to the jet radius).

For the jet viewing angle and for the speed we used the values given in
\cite{otterbein+1998}, that is: $\theta_\mathrm{j}=3^\circ$ , and a Lorentz
factor $\gamma_\mathrm{j} = 12$, which corresponds to an apparent
speed of $\beta_\mathrm{app} = 10.7$. The apparent pattern speed was
measured by comparing ridge lines for different epochs at $15$\,GHz in the
images of the MOJAVE monitoring program. This calculation was made using only the highest S/N region at
1--4\,mas distances from the jet origin. The measured speed was
$w_\mathrm{app} = 0.35 \pm 0.25 c$. With these quantities, the Mach
number $M_\mathrm{j}$ and the ratio of the jet to the ambient density $\eta$ can be
calculated as \citep[\textit{e.g.},][]{hardee2000}

\begin{equation}
    M_\mathrm{j} = \frac{\lambda^*(1-\beta_w\cos{\theta_\mathrm{j}})}{8R_\mathrm{j}\gamma_\mathrm{j}(1-\beta_w/\beta_\mathrm{j})\sin{\theta_\mathrm{j}}}\,,
\end{equation}
\begin{equation}
    \eta =\frac{M_\mathrm{j}^2}{M_\mathrm{x}^2},    
\end{equation}
where the external jet Mach number, $M_\mathrm{x}$, and the intrinsic jet pattern speeds are calculated with
\begin{equation}
    M_\mathrm{x} = \frac{\lambda^*\beta_\mathrm{j}(1-\beta_w\cos{\theta_\mathrm{j}})}{8R_\mathrm{j}\beta_w\sin{\theta_\mathrm{j}}}\,,
\end{equation}
\begin{equation}
\beta_w = \frac{w_\mathrm{app}}{\sin{\theta_\mathrm{j}}+w_\mathrm{app}\cos{\theta_\mathrm{j}}}\,, 
\end{equation}
and
\begin{equation}
    \beta_\mathrm{j} = \frac{\beta_\mathrm{app}}{\sin{\theta_\mathrm{j}}+\beta_\mathrm{app}\cos{\theta_\mathrm{j}}}.
\end{equation}
This yields a Mach number of $M_\mathrm{j} = 12 \pm 3$, in contrast to
the results given in \cite{lobanov+1998}, who reported a Mach number of
6 and a density ratio of $\eta = 0.33 \pm 0.08$. The ratio of the jet to the ambient
density we obtained here is close to
  unity, which may indicate that the jet is surrounded by a very
diluted cocoon. This restriction could be alleviated if the pattern
speed we used represents an upper limit because the different modes travel
at different speeds; the shorter wavelength modes travel
faster. Taking into account that the pattern speed was calculated in a
rather small region, it therefore probably corresponds to a mode with
a small wavelength, which is difficult to measure. The effect of the
speed on the ratio of the jet to the ambient density is also relevant. If we look
at different values within the range allowed by the error, for example
assuming a pattern speed of $w_\mathrm{app} = 0.10 c$, the
ratio of the jet to the ambient density changes by an order of magnitude, to
$\eta \sim 0.02$. Nevertheless, the jet Mach number remains
unchanged. Another important remark is that the approximation used
before is only valid for cases where $M_\mathrm{j} \gg 1$, which is
verified by our result, and in the scenario where a contact
discontinuity (or a narrow shear-layer) separates the jet and the
ambient medium. The possibility of a shear layer playing a
  role in the jet in S5\,0836+710 has been suggested in earlier works
  \citep{perucho+2007b,perucho+2011} after solving the linear
  stability problem for sheared jets, using a set of jet parameters
  given by \citet{lobanov+1998}. Arising naturally in
    spine-sheath scenarios, such a shear layer will also affect the
    growth rates and wavelengths of the KH instability modes
    \citep[see][]{mizuno+2007,hardee2007}.

\section{Summary}
\label{sc:summary}

Multiband VLBI observations of S5\,0836+710 with {\em RadioAstron}
provided images of the jet with an unprecedentedly high angular
resolution, reaching down to 15 microarcseconds at 22\,GHz, which
corresponds to a linear scale of 0.13~pc. The radio source S5\,0836+710 was observed with a ground- and space-VLBI array at the frequencies of 1.6 GHz on 24 October 2013, and 5
and 22 GHz on 10 January 2014.  The source was observed with tracks
of 16.5 hours at 1.6\,GHz and over a 30-hour period (with a 4-hour gap)
at 5 and 22\,GHz.
  
Non-standard procedures were needed to detect interferometric
fringes with the space antenna; the resulting residual delay solutions
for the 5 GHz {\em RadioAstron} image are shown in
Fig.~\ref{fg:fringe-sols}. They show a weak time-dependence due to
the acceleration of the space antenna.  The source was detected on
baselines as long as 10 Earth diameters at 1.6\,GHz, and 12 Earth
diameters at 5\,GHz and 22\,GHz.

Hybrid imaging was performed for the ground- and space-array data.
The latter yield resolutions of 3 to 10 times the ground resolution,
reaching scales of $\sim 15\,\mu$as ($\sim 0.12$\,pc) at 22\,GHz. The
full-resolution {\em RadioAstron} images show a much richer structural
detail than the ground-array images at the same
frequencies. At 5\,GHz and 22\,GHz, the jet is transversely resolved
in the {\em RadioAstron} images, revealing a bent and asymmetric
pattern embedded into the flow.

The observed brightness temperature of the jet is estimated for the
three frequencies and is listed in Table~\ref{tb:modelfit}.  These
estimates indicate that the highest brightness temperatures for each
frequency are $3.3 \times 10^{12}$\,K, $1.7 \times 10^{13}$\,K, and
$3.8 \times 10^{12}$\,K for 1.6\,GHz, 5\,GHz, and 22\,GHz,
respectively. These estimates agree well with the minimum
brightness temperature estimated from the visibility data, using the
longest $10$ $\%$ baselines. All of the estimates imply
$T_\mathrm{b}\ge 10^{13}$\,K in the reference frame of the source and
would require Doppler factors of up to $\sim 300$ in order to
reconcile them with the inverse Compton limit.

The internal structure of the jet was studied by analyzing the ridge line of
the jet. For the lowest frequency of 1.6\,GHz, the location of the
ridge line observed in the ground-array and space-VLBI images were
obtained by fitting transverse brightness profiles with a single-Gaussian component and recording the position of its peak with respect
to the overall jet axis oriented at a position angle of
$-$162$^\circ$. The ridge lines were represented with a simple model as
the sum of multiple oscillatory terms. The parameters of these
oscillatory modes are modeled and explained in the framework of
a Kelvin-Helmholtz instability that develops in the flow. The wavelengths
of the oscillations are found to be similar to those reported in
previous studies of the source. These oscillations can be interpreted
as resulting from the helical modes of the instability developing in
the jet. Based on this interpretation, an estimate of the jet Mach
number and the ratio of the jet to  the ambient density is obtained of $\sim 12$ and
$\sim 0.33,$ respectively. This estimate can be further verified and
refined using a more detailed numerical analysis of the jet
stability.

\section*{Acknowledgments}

L.V.G. is a member of the International Max Planck Research School
(IMPRS) for Astronomy and Astrophysics at the Universities of Bonn and
Cologne. The {\em RadioAstron} project is led by the Astro Space
Center of the Lebedev Physical Institute of the Russian Academy of
Sciences and the Lavochkin Scientific and Production Association under
a contract with the State Space Corporation ROSCOSMOS, in
collaboration with partner organizations in Russia and other
countries.  This research is based on observations correlated at the
Bonn Correlator, jointly operated by the Max Planck Institute for
Radio Astronomy (MPIfR), and the Federal Agency for Cartography and
Geodesy (BKG).  The European VLBI Network is a joint facility of
European, Chinese, South African and other radio astronomy institutes
funded by their national research councils. The National Radio
Astronomy Observatory is a facility of the National Science Foundation
operated under cooperative agreement by Associated Universities, Inc.
Thanks to Phillip Edwards and Alan Roy for the useful comments about
the paper. M.P. has been supported by the Spanish Ministerio de
Econom\'{\i}a y Competitividad (grants AYA2015-66899-C2-1-P and
AYA2016-77237-C3-3-P) and the Generalitat Valenciana (grant
PROMETEOII/2014/069).  This work was partially supported by the COST
Action MP0904 Black Holes in a Violent Universe. G.B. acknowledges
financial support under the INTEGRAL ASI-INAF agreement 2013-025-R.1.
T.S. was supported by the Academy of Finland projects 274477, 284495,
and 312496. I.A. acknowledges support by a Ram\'on y Cajal grant of
the Ministerio de Econom\'{i}a, Industria y Competitividad (MINECO) of
Spain. The research at the IAA-CSIC was partly supported by the MINECO
through grants AYA2016-80889-P, AYA2013-40825-P, and
AYA2010-14844. Y.Y.K. was supported in part by the government of the
Russian Federation (agreement 05.Y09.21.0018) and the Alexander von
Humboldt Foundation.

\bibliographystyle{aa}
\bibliography{ra0836}


\begin{appendix}

 \section{Auxiliary plots: {\em uv} coverage, visibility amplitudes, hybrid images, and jet ridge lines.}
 \label{ap:images}
 
\begin{figure}[ht]
  \centering
  \includegraphics[height=0.39\textheight, trim = 35mm 0mm 35mm 0mm, clip=true]{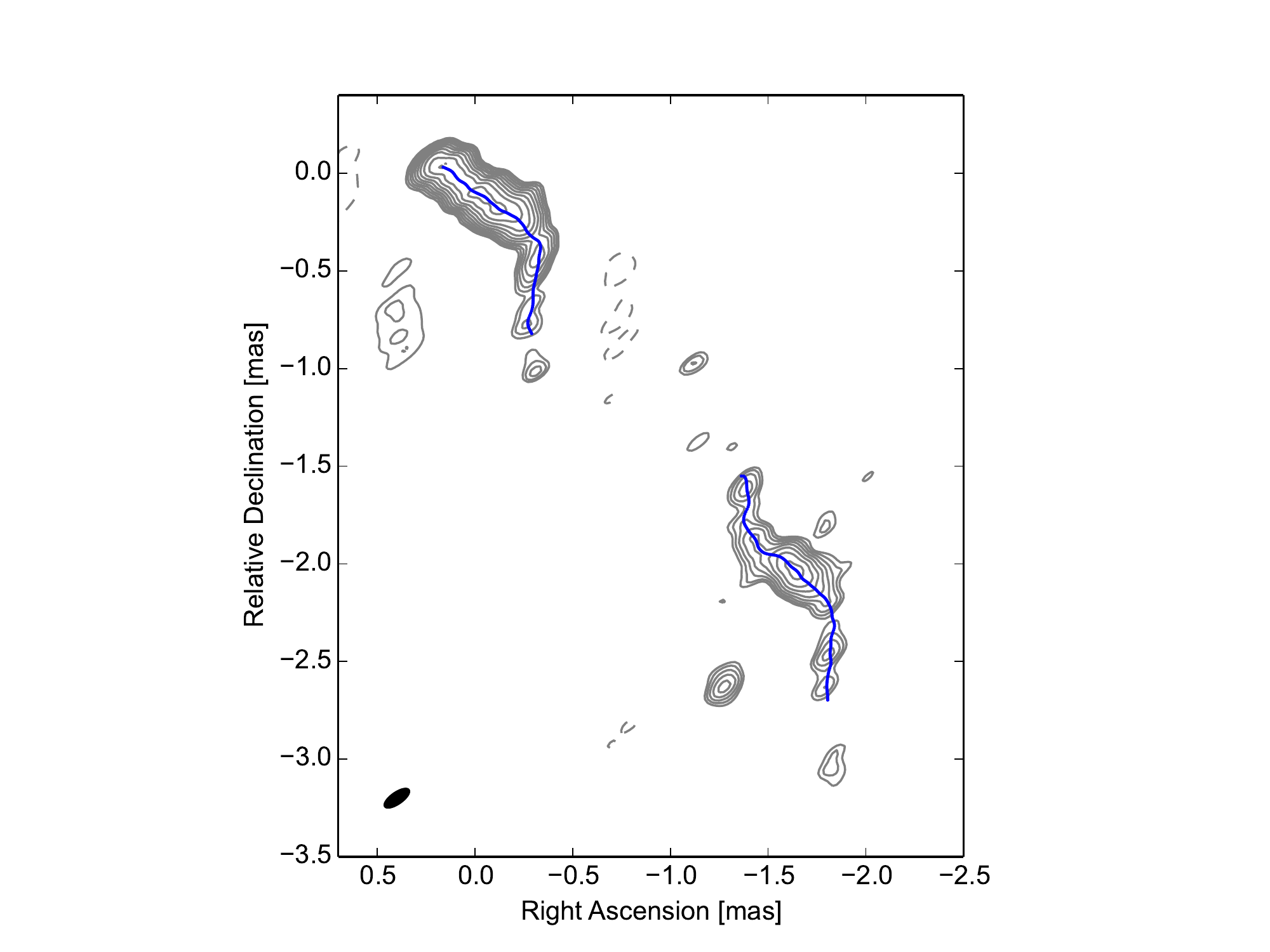}
  \caption{Ridge line in the {\em RadioAstron} image of the jet in S5\,0836$+$710 at 
5\,GHz shown in Fig.~\ref{fg:0836-5-22-ghz-maps}.
The contour levels are drawn at ($-$1, 1,
    $\sqrt{2}$, 2, etc. ) times 7.0~mJy/beam. Image parameters are given in Table~\ref{tb:mapspar}. A discussion of the jet ridge lines is presented in
    Sect.~\ref{sc:ridgeLineOscillations}.}
\label{fg:0836-5-sridge} 
\end{figure}

 This appendix presents auxiliary information and images obtained
 from the {\em RadioAstron} observations. Ridge lines determined
 from {\em RadioAstron} images at 5\,GHz and 22\,GHz are shown in
 Figs.~\ref{fg:0836-5-sridge}--\ref{fg:0836-22-sridge}. The ground
-array images and respective ridge lines at 43, 22, 15, and 5\,GHz are
 presented in
 Figs.~\ref{fg:0836-43-gridge}--\ref{fg:0836-5-gridge}. The {\em
   uv} coverage of the accompanying ground-array observations at 15
 and 43\,GHz are shown in
 Figs.~\ref{fg:0836-15-uvplot}--\ref{fg:0836-43-uvplot}. Radial
 distributions of the visibility amplitudes measured in the {\em
   RadioAstron} observations at 1.6, 5, and 22\,GHz are given in
 Fig.~\ref{fg:0836-radplot}.

\vfill\eject
\newpage

\setcounter{figure}{1}
\begin{figure}[ht]
  \centering
  \includegraphics[height=0.39\textheight, trim = 30mm 0mm 35mm 0mm, clip=true]{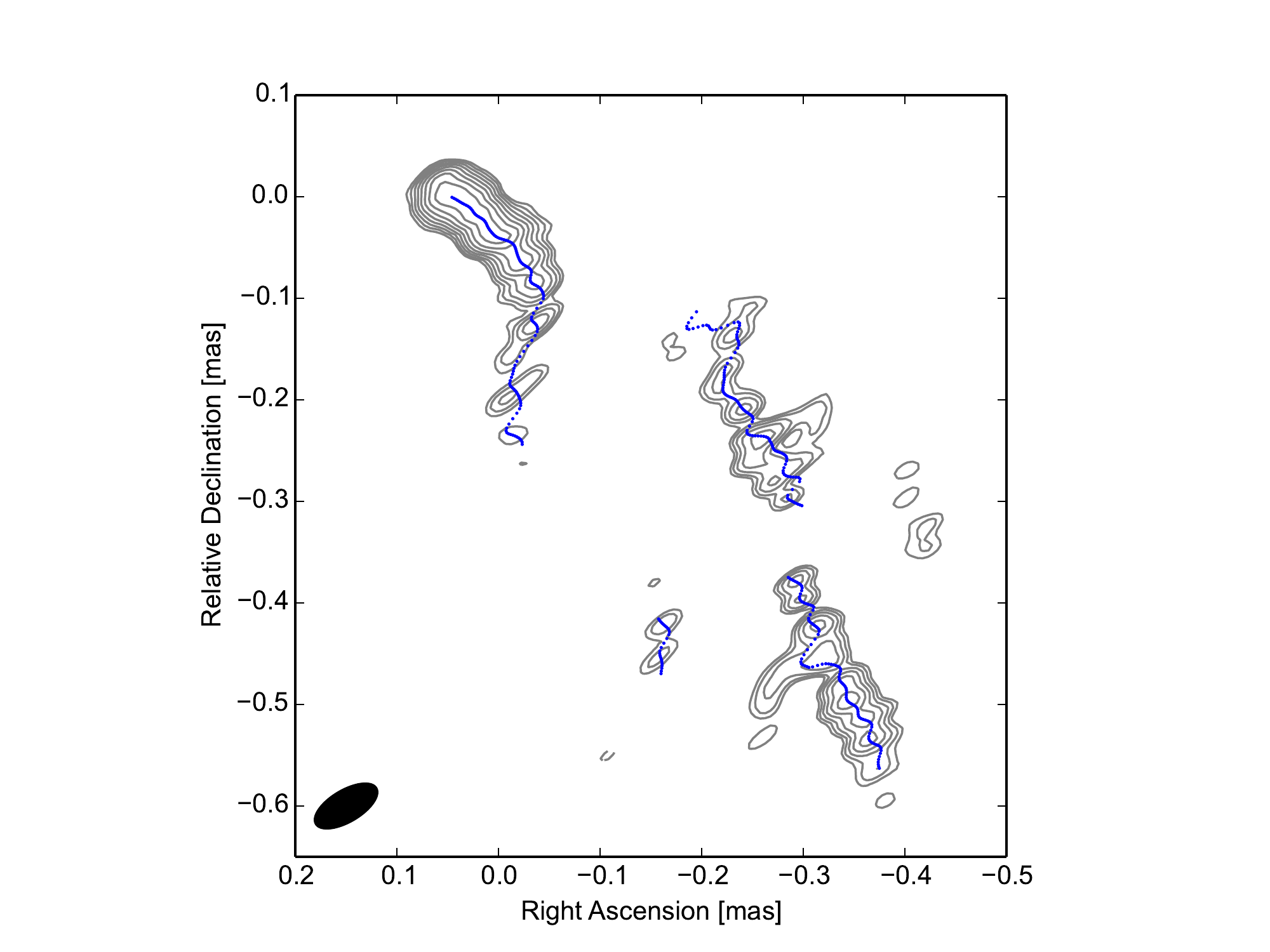}
\caption{Ridge line in the {\em RadioAstron} image of the jet in
  S5\,0836$+$710 at 22\,GHz shown in Fig.~\ref{fg:0836-5-22-ghz-maps}.  The
  contour levels are drawn at ($-$1, 1, $\sqrt{2}$, 2, etc. ) times 5.0
  mJy/beam. Image parameters are given in
  Table~\ref{tb:mapspar}. A discussion of the jet ridge lines is
  presented in Sect.~\ref{sc:ridgeLineOscillations}.}
\label{fg:0836-22-sridge} 
\end{figure}

\vfill\eject
\newpage

\setcounter{figure}{2}
\begin{figure}[h!]
  \centering
  \includegraphics[height=0.35\textheight]{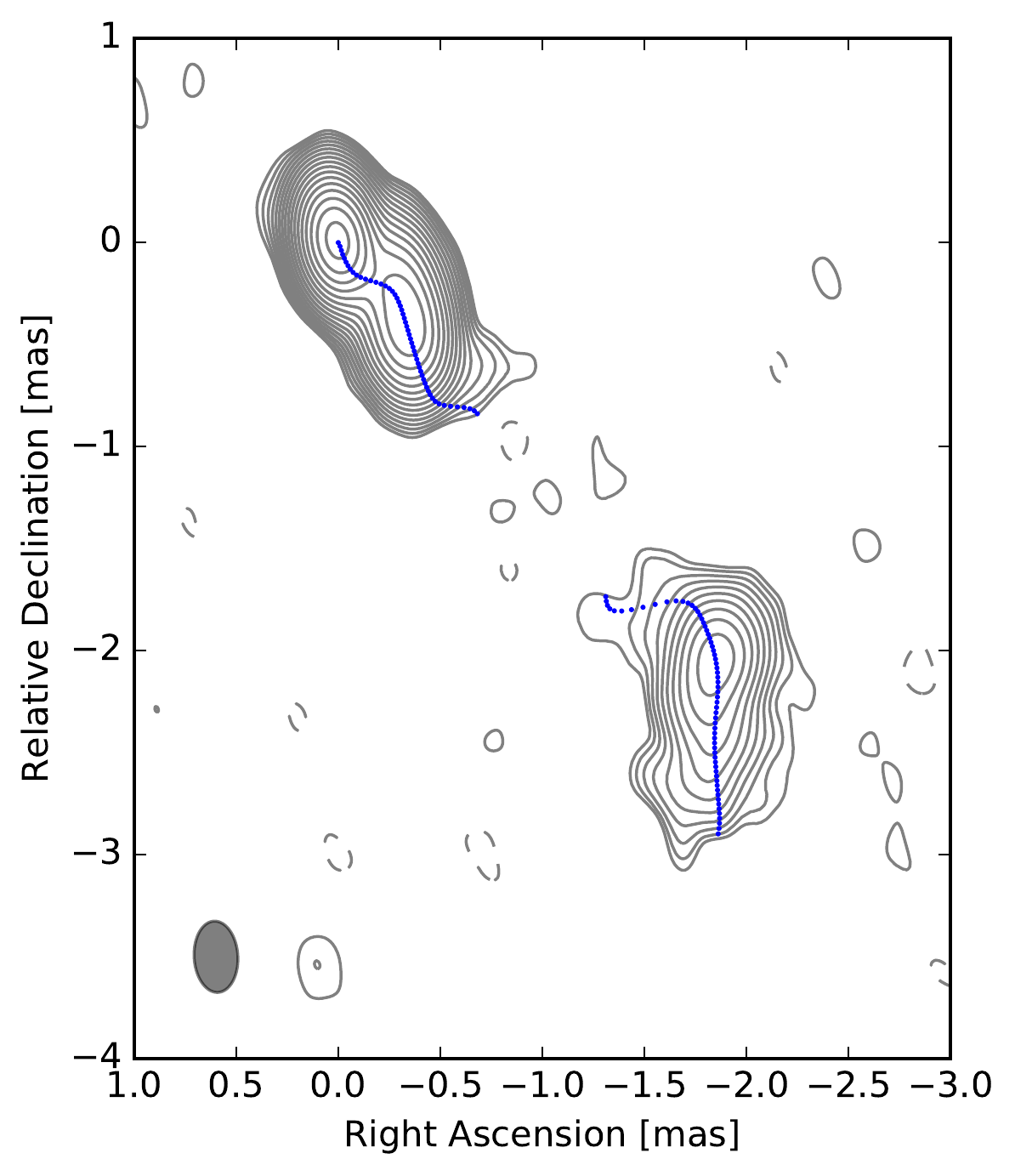}
  \caption{Image of the jet in S5\,0836$+$710 made from the ground-VLBI
    data at 43\,GHz.  The contour levels are drawn at ($-$1, 1,
    $\sqrt{2}$, 2, etc. ) times 1.8~mJy/beam. The curved blue line
    denotes the ridge line of the jet we derived that is discussed in
    Sect.~\ref{sc:ridgeLineOscillations}.}
\label{fg:0836-43-gridge} 
\end{figure}

\setcounter{figure}{4}
\begin{figure}[hb]
  \centering
  \includegraphics[height=0.40\textheight]{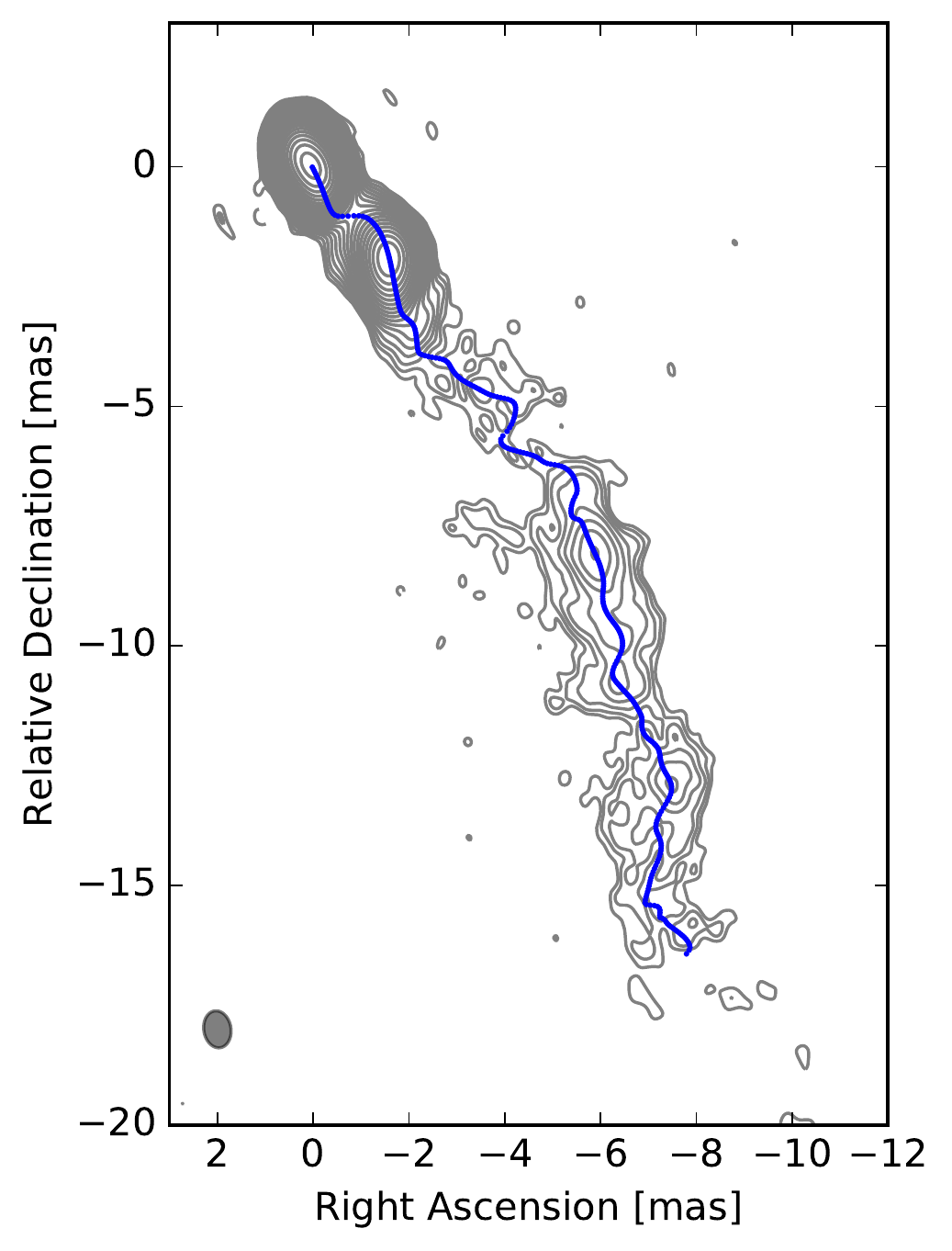}
  \caption{Image of the jet in S5\,0836$+$710 made from the ground-VLBI data at 15\,GHz. 
  The contour levels are drawn at ($-$1, 1, $\sqrt{2}$, 2, etc. ) times 0.75~mJy/beam. The curved blue line denotes the ridge line of the jet we derived that is discussed in Sect.~\ref{sc:ridgeLineOscillations}.}
\label{fg:0836-15-gridge} 
\end{figure}

\setcounter{figure}{3}
\begin{figure}[ht]
  \centering
  \includegraphics[height=0.35\textheight]{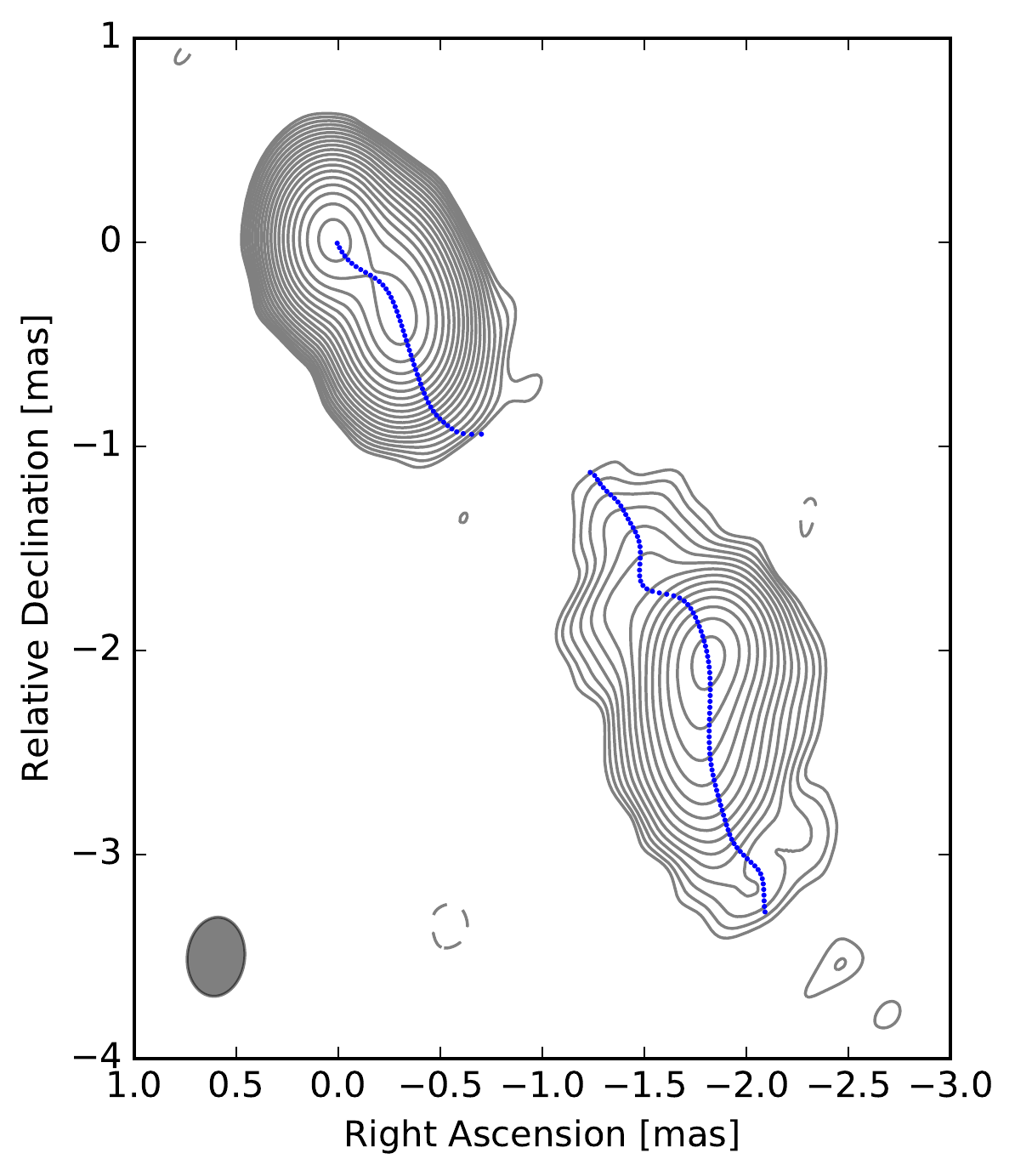}
  \caption{Image of the jet in S5\,0836$+$710 made from the ground-VLBI data at 22\,GHz.
  The contour levels are drawn at ($-$1, 1, $\sqrt{2}$, 2, etc. ) times 1.0~mJy/beam. The curved blue line denotes the ridge line of the jet we derived that is discussed in Sect.~\ref{sc:ridgeLineOscillations}.}
\label{fg:0836-22-gridge} 
\end{figure}

\setcounter{figure}{5}
 \begin{figure}[htb]
  \centering
  \includegraphics[height=0.42\textheight]{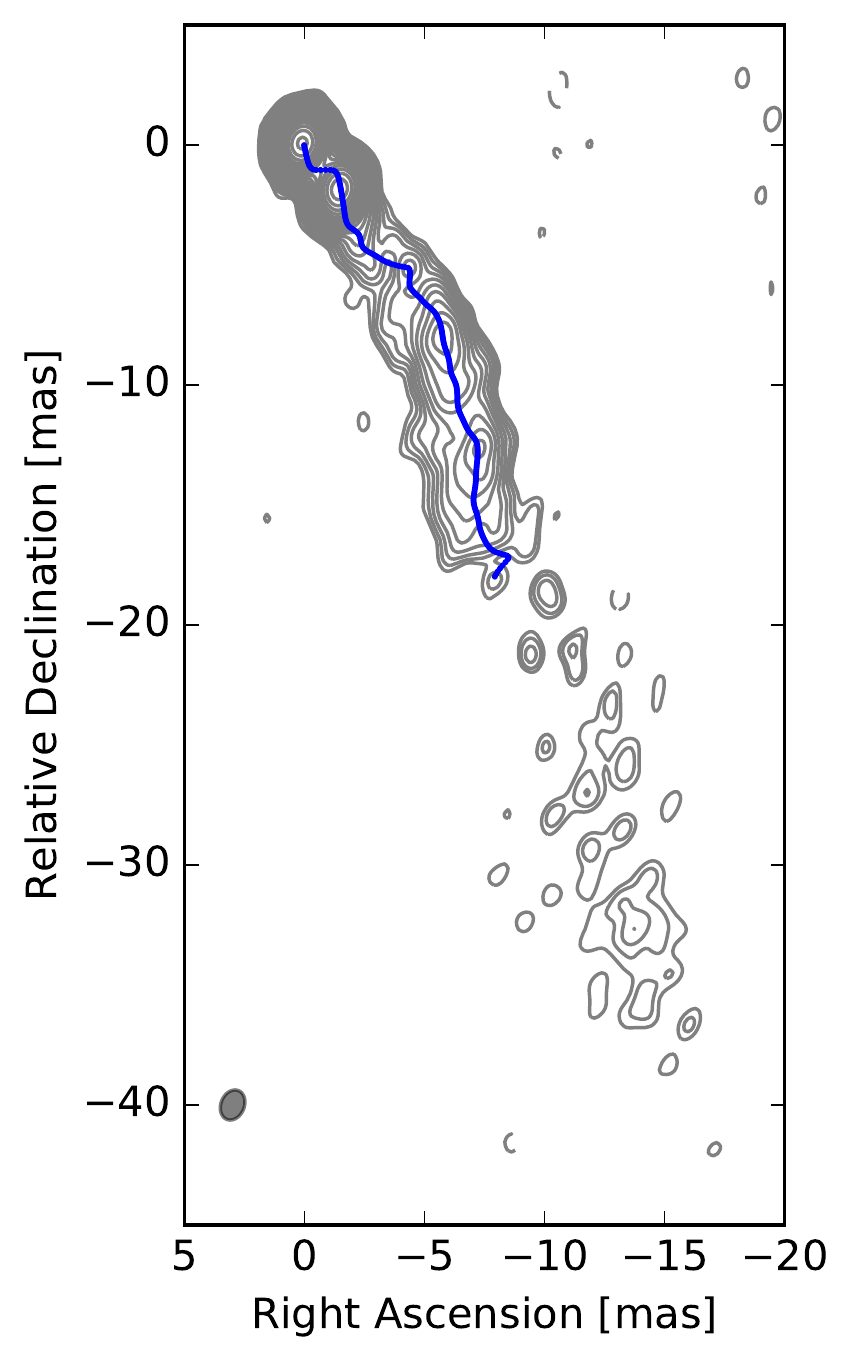}
  \caption{Image of the jet in S5\,0836$+$710 made from the ground-VLBI data at 5\,GHz.
  The contour levels are drawn at ($-$1, 1, $\sqrt{2}$, 2, etc. ) times 1.5~mJy/beam. The curved blue line denotes the ridge line of the jet we derived that is discussed in Sect.~\ref{sc:ridgeLineOscillations}.}
\label{fg:0836-5-gridge} 
\end{figure}

\setcounter{figure}{6}
 \begin{figure}[ht]
  \centering
\centerline{\includegraphics[width=0.45\textwidth,angle=0,trim=12mm 42mm 25mm 52mm,clip=true]{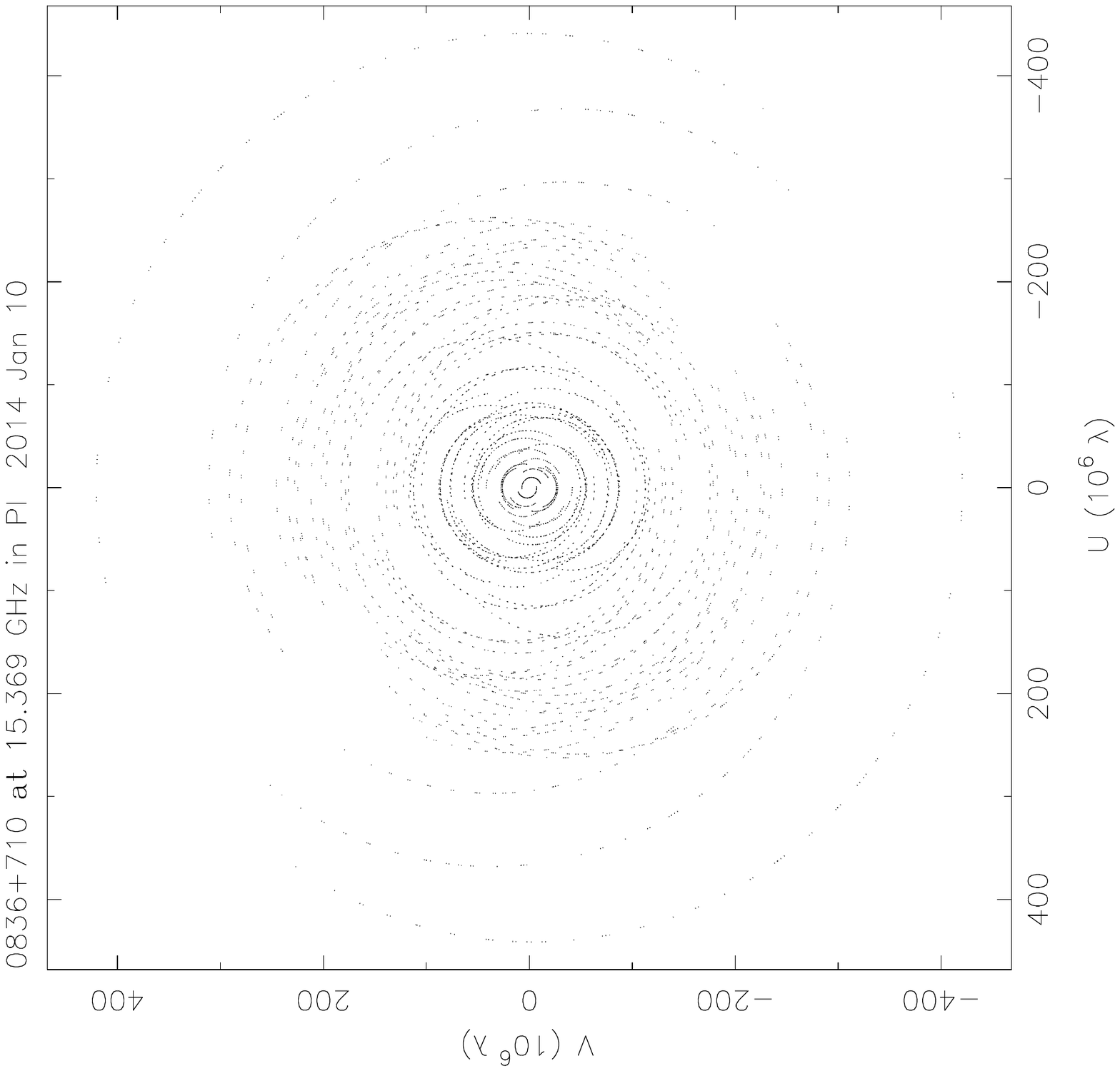}}
\caption{Coverage of the Fourier domain ({\em uv} coverage) of the ground-array observations of S5\,0836$+$710 at 15\,GHz plotted in units of M$\lambda$.}
\label{fg:0836-15-uvplot}
\end{figure}

\setcounter{figure}{7}
 \begin{figure}[ht]
  \centering
\centerline{\includegraphics[width=0.45\textwidth,angle=0,trim=12mm 42mm 25mm 52mm,clip=true]{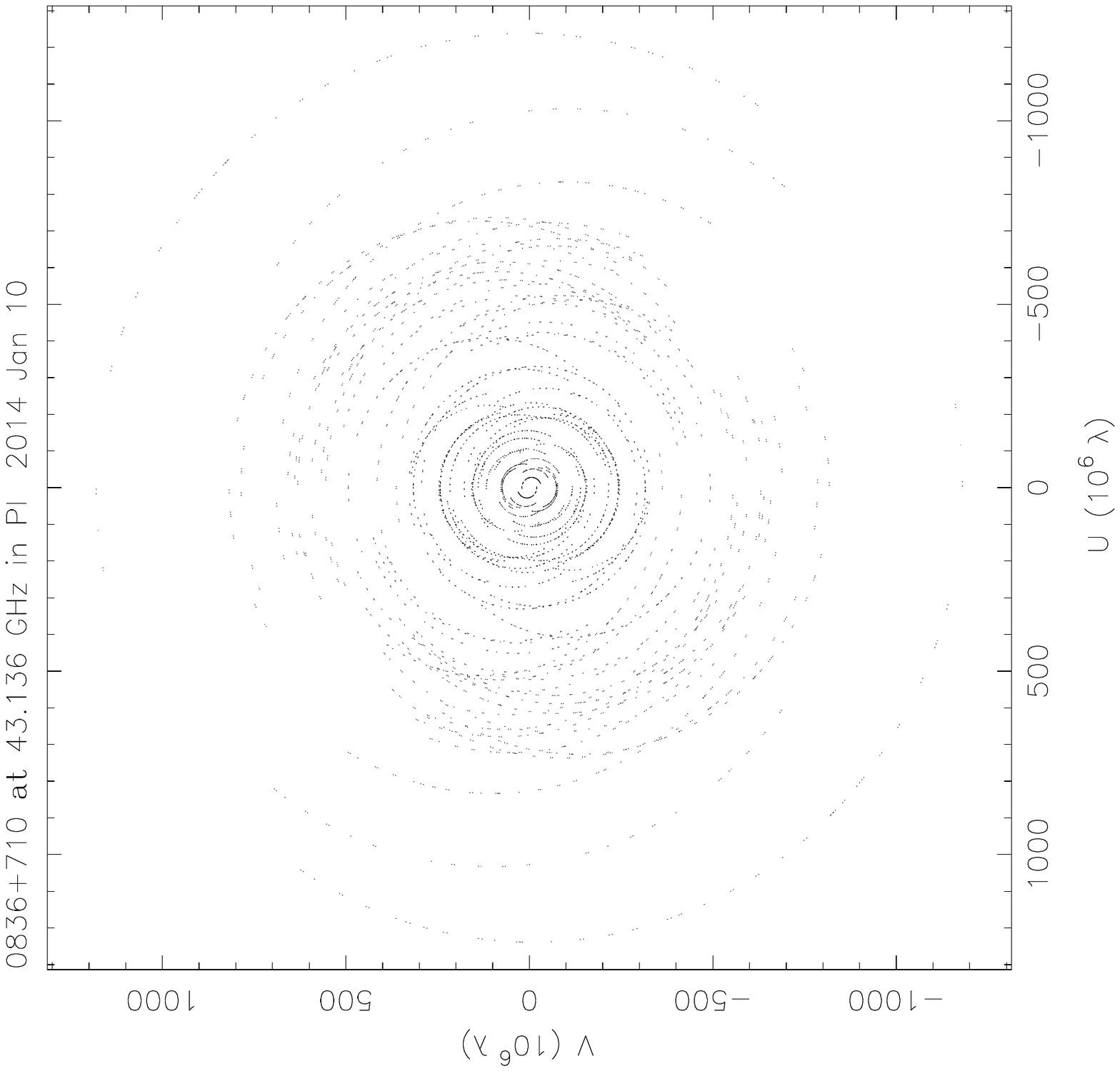}}
\caption{Coverage of the Fourier domain ({\em uv} coverage) of the ground-array observations of S5\,0836$+$710 at 43\,GHz plotted in units of M$\lambda$.}
\label{fg:0836-43-uvplot}
\end{figure}

\vfill
\newpage

\begin{figure}[hb!]
\centerline{\includegraphics[width=0.27\textheight,angle=-90,trim=62 26 34 60,clip=true]{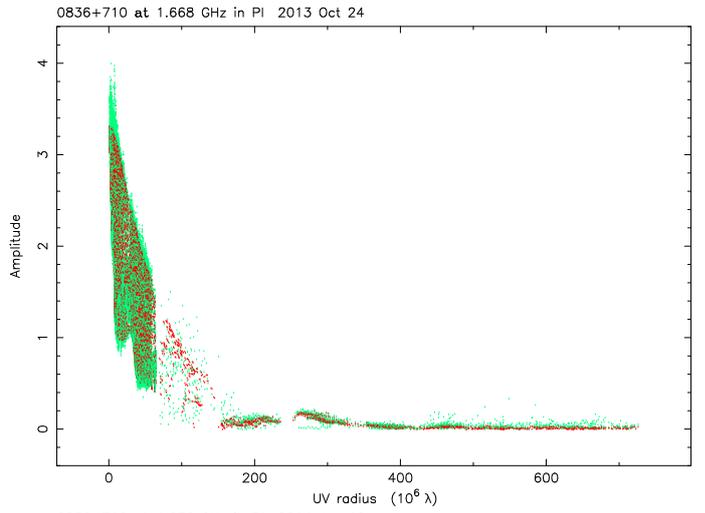}}
\centerline{\includegraphics[width=0.27\textheight,angle=-90,trim=62 26 34 60,clip=true]{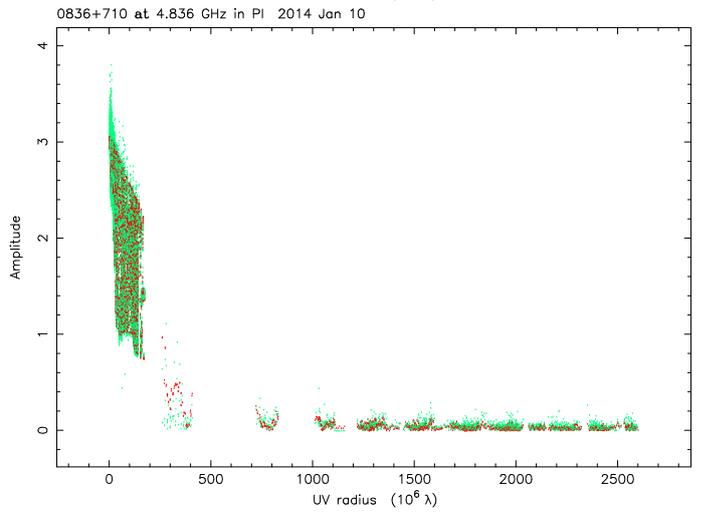}}
\centerline{\includegraphics[width=0.27\textheight,angle=-90,trim=62 26 34 60,clip=true]{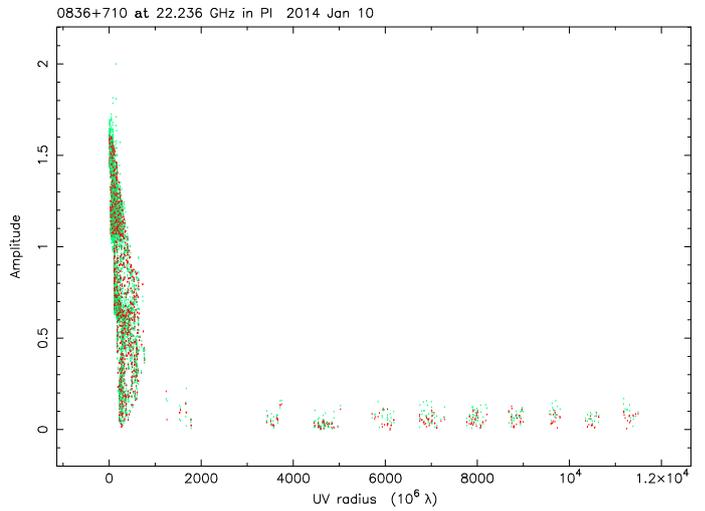}}
\caption{Radial distributions of visibility amplitudes (orange) from the {\em RadioAstron} observations of S5\,0836$+$710 at 1.6\,GHz (top), 5\,GHz (middle), and
  22\,GHz (bottom) as a function of {\em uv} distance, overplotted
  with the respective CLEAN models (red) of the source structure. }
\label{fg:0836-radplot} 
\end{figure}

\end{appendix}

\end{document}